\documentclass{article}
\usepackage{cite}
\usepackage{amsfonts}
\usepackage{amssymb}
\usepackage{amsmath}
\usepackage{graphicx}
\usepackage{slashed}
\usepackage{setspace}

\usepackage{color}
\usepackage{cancel}

\definecolor{darkred}{rgb}{0.5,0,0}
\definecolor{darkblue}{rgb}{0,0,0.4}
\definecolor{darkgray}{rgb}{0.3,0.3,0.3}
\definecolor{darkgreen}{rgb}{0.,0.23,0.1}

\def\hp{\mathord{+}}
\def\hm{\mathord{-}}

\newcommand{\cac}{{\cal C}}
\def \tk{\tilde k}
\newcommand{\sqNMHV}{{\rm \sqrt{N}MHV}}
\newcommand{\Z}{\mathbb{Z}}

\def\half{{1\over 2}}

\newcommand{\reef}[1]{(\ref{#1})}

\newcommand{\dash}{\text{-}}
\newcommand{\pdan}{{m}}

\newcommand{\ca}{{\cal A}}

\newcommand{\cn}{{\cal N}}
\newcommand{\cm}{{\cal M}}

 \newcommand{\co}{B^{\cdots}}

\newcommand{\be}{\begin{equation}}
\newcommand{\ee}{\end{equation}}

\def\be{\begin{equation}}
\def\ee{\end{equation}}
\def\bea{\begin{eqnarray}}
\def\eea{\end{eqnarray}}
\def\ba{\begin{array}}
\def\ea{\end{array}}
\def\bd{\begin{displaymath}}
\def\ed{\end{displaymath}}

\def\ie{{\it i.e.~}}

\def\a{\alpha}
\def\b{\beta}

\def\d{\delta}
\def\e{\epsilon}           
  
\def\g{\gamma}
\def\h{\eta}

\def\l{\lambda}
\def\m{\mu}

\def\s{\sigma}                                   

\def\D{\Delta}

\def\pa{\partial}                              
\def\>{\rangle} 
\def\<{\langle} 
\def\Dsl{D \hskip-.6em \raise1pt\hbox{$ / $ } }
\def\to{\rightarrow}

\def\pa{\partial}

\def\lab{\label}

\def\tk{\tilde{k}}
\def\cc{{\cal C}}

\newcommand{\tQd}{\widetilde Q}
\newcommand{\eps}{\epsilon}
\newcommand{\lra}{\leftrightarrow}

\def\tQ{\widetilde{Q}}

\begin{document}

\setstretch{1.05}

\begin{titlepage}

\begin{flushright}
MCTP-10-56\\
MIT-CTP-4197  \\
PUPT-2361
\end{flushright}
\vspace{.5cm}

\centerline{\bf \Large SUSY Ward identities, Superamplitudes, and Counterterms}

\begin{center}
{\bf Henriette Elvang${}^{a,b}$,
Daniel Z.~Freedman${}^{c,d}$, Michael Kiermaier$^{e}$} \\
\vspace{0.7cm}
{${}^{a}${\it Michigan Center for Theoretical Physics, Randall Laboratory of Physics}\\
{\it University of Michigan, Ann Arbor, MI 48109, USA}}\\[2mm]
{${}^{b}${\it School of Natural Sciences, Institute for Advanced Study, Princeton, NJ 08540, USA}}\\[2mm]
{${}^{c}${\it Department of Mathematics, Massachusetts Institute of Technology, Cambridge, MA 02139, USA}}\\[2mm]
{${}^{d}${\it Center for Theoretical Physics, Massachusetts Institute of Technology, Cambridge, MA 02139, USA}}\\[2mm]
{${}^{e}${\it Joseph Henry Laboratories, Princeton University, Princeton, NJ 08544, USA}}\\[3mm]
{\small \tt  elvang@umich.edu,
 dzf@math.mit.edu, mkiermai@princeton.edu}
\end{center}

\begin{abstract}
Ward identities of SUSY and R-symmetry relate $n$-point amplitudes in supersymmetric theories. We review recent work in which these Ward identities are solved in $\cn =4$ SYM and $\cn=8$ supergravity. The solution, valid at both tree and loop level, expresses any N$^K$MHV superamplitude in terms of a basis of ordinary amplitudes. Basis amplitudes are classified by semi-standard tableaux of rectangular $\cn$-by-$K$ Young diagrams.
The SUSY Ward identities also impose
constraints on the matrix elements of candidate ultraviolet counterterms in $\cn=8$ supergravity,
and they
can be studied using superamplitude basis expansions. This leads to a novel and
quite
comprehensive matrix element approach to counterterms, which we also review.

This article is an invited review for a special issue of  Journal of Physics A devoted to  ``Scattering Amplitudes in Gauge Theories''.
\end{abstract}

\end{titlepage}

\vspace{4mm}
\setstretch{0.3}
\setcounter{tocdepth}{2}
\tableofcontents
\setstretch{1.05}

\newpage

\setcounter{equation}{0}
\section{Introduction}

Supersymmetry and R-symmetry Ward identities impose linear relations among
individual amplitudes in supersymmetric theories. The first question addressed in this review is how to solve the Ward identities
in $\cn=4$ SYM and $\cn=8$ supergravity
and specify a basis of amplitudes  that determines all others in the same  N$^K$MHV class.  For MHV amplitudes, the answer is simple: any one amplitude determines the entire class.  However,  for $K \ge 1$ very little information was known until recently. We review the results of~\cite{Elvang:2009wd}, where the SUSY and R-symmetry Ward identities were solved to give an expansion of the general N$^K$MHV superamplitude in terms of a minimal basis of component amplitudes that are independent under these Ward identities.
 In the second part of this review, we apply this expansion to the analysis of potential counterterms in $\cn=8$ supergravity~\cite{Elvang:2010jv}. Imposing the additional requirement of locality on the manifestly SUSY and R-invariant expansion of superamplitudes is at the heart of this {\em matrix-element approach to counterterms}. Just as recursion relations focus on on-shell scattering amplitudes instead of Lagrangians, the center of attention is shifted from counterterm operators to their matrix elements.

The first approach to SUSY Ward identities for on-shell amplitudes was the 1977 work of Grisaru and Pendleton \cite{Grisaru:1977px} (see also \cite{Grisaru:1976vm, Parke:1985pn}). They discussed the structure of these identities and solved them for 6-point NMHV amplitudes in $\cn =1$ SUSY.  Six basis amplitudes were needed to determine all 60 NMHV amplitudes.\footnote{The solution of \cite{Grisaru:1977px} was rederived using modern spinor-helicity methods in \cite{Bianchi:2008pu}.}

A general solution to the $\cn =4$ and $\cn = 8$ Ward identities was recently presented in \cite{Elvang:2009wd} and will be reviewed in
Sec.~\ref{secNMHV}-\ref{n8samps}
below. The solution exploits the properties of superamplitudes which compactly encode all individual $n$-point amplitudes at each N$^K$MHV level.  The Ward identities can be elegantly imposed as constraints on the superamplitudes which are then expressed as sums  of simple manifestly SUSY- and
R-invariant Grassmann polynomials, each multiplied by an ordinary amplitude. This set of ordinary amplitudes comprise a basis for the superamplitude. Only the Ward identities of non-anomalous  Poincar\'e SUSY and $SU(\cn)_R$ symmetry are  used, so the results apply both to tree and loop amplitudes.   The dual conformal and Yangian symmetries of the $\cn =4$ theory are important and have led to much new information about planar amplitudes of the theory.\footnote{
See~\cite{Bargheer:2011mm} for a review in this issue.} 
Those symmetries were
not included in the analysis of \cite{Elvang:2009wd}, so that the results are valid for both planar and nonplanar amplitudes of $\cn =4$ SYM and also for $\cn=8$  supergravity.

Let us
provide a preview of the structure of superamplitudes and their basis expansion with details discussed in Sec.~\ref{secsusyR}-\ref{n8samps}.
Superamplitudes~\cite{Nair:1988bq,Witten:2003nn, Georgiou:2004by}  ${\cal A}_n$  are generating functions
 for ordinary amplitudes whose bookkeeping Grassmann variables $\eta_{ia}$
are labeled by particle number  $i=1,\ldots, n$ and by  the $SU(\cn)_R$-symmetry index~$a=1,\ldots, \cn$. At level N$^K$MHV,
 the superamplitudes  are Grassmann  polymomials of order $\cn(K+2)$. Their coefficients are the actual scattering amplitudes.

Supercharges ${\ Q}^a$ and ${ \tQ}_a$
defined by the simple expressions
\be   \lab{tqq}
{ \widetilde Q}_a = \sum_{j=1}^n\, | j\>\, \eta_{ja}\,, \qquad\quad
{ Q}^a = \sum_{j=1}^n\,| j]\, \frac{\pa}{\pa\eta_{ja}} \,
\ee
act directly on the superamplitudes, giving the Ward identities
\be \lab{susyward}
{\ \widetilde Q}_a \, {\cal A}_n =0, \qquad\qquad { Q}^a {\cal A}_n =0\,.
\ee
It is these SUSY Ward identities combined with important constraints due to R-symmetry which are solved in \cite{Elvang:2009wd}.  The solutions derived for superamplitudes take the schematic form
\be
  \lab{cartoon}
  \ca_{n}^\text{N$^K$MHV} ~=~
  \sum_I A_I \, Z_I \, .
\ee
The index $I$ enumerates the set of independent SUSY and R-symmetry invariant Grassmann polynomials $Z_I$ of degree $\cn(K+2)$.
They are constructed from two simple and familiar ingredients,
 which are explained in more detail below.
First,  each $Z_I$ contains a factor of the well known Grassmann delta-function,
 $\d^{(2\cn)}(\tQd)$,
which expresses the conservation of $\widetilde Q_a$. It is a degree $2\cn$ polynomial which is
annihilated by both $Q^a$ and $\widetilde Q_a$.  The
 other ingredient is that each $Z_I$ contains $\cn K$ factors of the first-order polynomial
\begin{equation}\label{pdanIntro}
    \pdan_{ijk,a} \equiv [ij] \h_{ka} +[jk] \h_{ia} +[ki] \h_{ja}\,,
\end{equation}
in which  $i,j,k$ label three external lines of the $n$-point amplitude.  Every polynomial $\pdan_{ijk,a}$  is annihilated by $Q^a$. The polynomial \reef{pdanIntro}
is the essential element of the well-known 3-point anti-MHV superamplitude.

The basis amplitudes $A_I$ in \reef{cartoon}
are matrix elements for specific particle processes within each N$^K$MHV sector.  Finding the basis can be formulated as a group theoretic problem, and it has a neat solution.  The number of amplitudes in the  basis is the dimension of the irreducible representation of $SU(n-4)$ corresponding to a rectangular Young diagram with $K$ rows and  $\cn$ columns! The independent amplitudes are precisely labeled by the semi-standard tableaux of this Young diagram.

As an example, consider the $6$-point NMHV amplitude ${\cal A}^{\rm NMHV}_6$ in $\cn=4$ SYM.
There are 5 basis amplitudes which can be chosen to be  the 6-point matrix elements:
\begin{equation}
 \<-+++--\>\,,\quad \<\l^-\l^+++--\>\,,\quad
 \<s\,\bar s ++--\>\,,
 \quad\<\l^+\l^-++--\>\,,\quad\<+-++--\>\,.
\end{equation}
The last 4 particles in each amplitude are `standardized' by SUSY
to be gluons of positive and negative helicity. In the first two positions we  must allow any combination that leads to an NMHV amplitude, \ie pairs of gluons, gluinos, and scalars.  For $\cn=8$  supergravity,
 the analogous basis contains  9 basis amplitudes which we can again specify to contain  `standardized' gravitons as the last 4 particles and pairs of gravitions,  gravitinos, etc.
 on the first two lines.

Basis amplitudes containing four gluons `$++-\,-$' on four fixed lines are particularly convenient
to write down the superamplitude in closed form. Using a computer-based implementation of this superamplitude, however, one can choose any other set with the same number of linearly independent amplitudes. Linear independence, in this case, is best verified numerically. At the $6$-point NMHV level, for example, a suitable basis of $5$ linearly independent gauge theory amplitudes is the split-helicity
 gluon amplitude $\<+++---\>$ together with $4$ of its cyclic permutations,
specifically
\be
  \<+++---\>,~~~~
  \<-+++--\>,~~~~
  \<--+++-\>,~~~~
  \<---+++\>,~~~~
  \<+---++\>
  \,.
\ee
In $\cn=8$ supergravity, the pure graviton amplitude $M_{6}(+++---)$ together with $8$ permutations of its external lines  represents a suitable basis. It is striking that the basis of planar $\cn=4$ SYM ($\cn=8$ supergravity) at the $6$-point NMHV level reduces to momentum permutations of a single all-gluon (all-graviton) amplitude.

\medskip


The {\em second major topic} of this review is the application of the
basis expansions of superamplitudes to  candidate counterterms of the form
$\sqrt{-g}D^{2k}R^n$
 in the loop expansion of perturbative $\cn =8$ supergravity.
The matrix element method complements and extends other approaches
 to counterterms
which work with on-shell
superspace~\cite{superspaceCT,Drummond:2003ex,Drummond:2010fp}, information
from string theory \cite{stringCT,GVR,Eisenstein},  and  light-cone
superspace~\cite{lightCT}.
 The
 leading
 matrix elements of a potential counterterm
must be
 local and gauge invariant, and
this means that they are
polynomials in the spinor brackets
$\<i\,j\>,~[k\,l]$ associated with the external momenta.
Matrix elements of candidate counterterms at loop order $L$ are strongly
constrained by the overall scale
dimension and the helicities of their external particles.  In many cases
one can show quite simply that there are no
 local SUSY and R-invariant superamplitudes that
satisfy these constraints.   Then the corresponding operator
 is not supersymmetrizable and
cannot appear as an independent counterterm.
 On the other hand, when
the constraints are
satisfied,  the method explicitly constructs the  matrix elements of a
linearized
supersymmetric completion of the  operator.

In addition to SUSY and R-symmetry, the spontaneously broken $E_{7(7)}$  symmetry \cite{Cremmer:1979up} of $\cn =8$ supergravity gives additional constraints on counterterm matrix elements with external scalar particles.
 In particular, counterterm matrix elements must vanish in the single-soft scalar limit. These constraints were
analyzed to exclude the potential $3$-, $5$-, and $6$-loop counterterms $R^4$, $D^4R^4$, and $D^6R^4$ in the recent papers \cite{Elvang:2010kc, Beisert:2010jx} (see also~\cite{Brodel:2009hu,Eisenstein,Bossard:2010bd}),  which are reviewed in Sec.~\ref{secE77} below.

The net result of the matrix element approach to counterterms,
combined with the results of~\cite{Drummond:2010fp},
is that there are no admissible counterterms
in $\cn =8$ supergravity at loop order $L <7$. The method does not exclude counterterms at loop order $ L \ge 7$, but it shows that the only possible independent $L=7$ loop counterterm is $D^{8}R^4 + \dots$\,, whose leading matrix elements involve 4 external particles~\cite{Beisert:2010jx}; higher-point operators such as $D^4R^6$ (for which we present simple explicit superamplitude expressions in
Sec.~\ref{secexplicit})
and $R^8$ are compatible with SUSY and R-symmetry, but have non-vanishing single-soft scalar limits and thus violate continuous $E_{7(7)}$ symmetry~\cite{Beisert:2010jx} (see also~\cite{Kallosh:2010kk}). This implies that a computation of the $4$-point amplitude is sufficient to determine whether or not $\cn=8$ supergravity is finite at $7$-loop order.

In Sec.~\ref{s:lessRsym}, we discuss the structure of superamplitudes with reduced R symmetry. We focus on amplitudes that are invariant under an $SU(4)\!\times\!SU(4)$ subgroup of $SU(8)$; these are relevant both for the study of single-soft scalar limits in $\cn=8$ supergravity and for closed string tree amplitudes with massless external states in 4 dimensions.

\setcounter{equation}{0}
\section{SUSY Ward identities}\label{s:susywis}

Particle states of the $\cn=4$ and $\cn=8$ theories transform in anti-symmetric products of the fundamental representation
of the R-symmetry groups $SU(4)$ and $SU(8)$.
Thus the gluons, the 4 gluinos, and the 6 scalars of the $\cn=4$ theory can be described by
annihilation
 operators which carry anti-symmetrized upper indices:
\be  \label{n4ops}
B,~~B^a,~ ~B^{ab},~~B^{abc}, ~~B^{abcd}\,,
\ee
with $1\le a,b, \ldots \le 4$.  The tensor rank $r$ is related to the particle helicity $h$ by $2h = 2 -  r$. The 256 particle states of $\cn =8$ supergravity
 are
described analogously by annihilation operation operators of tensor rank $0\le r \le 8$.  Helicity and rank are related by $2h = 4 - r$.

We now discuss the S-matrix elements and Ward identities for the simpler $\cn=4$ theory.  The extension to $\cn =8$ is straightforward.  One can suppress indices and simply use $\co$ for any
annihilation operator from the set in \reef{n4ops}.  A generic $n$-point amplitude may then be denoted by
\be \lab{amp1}
A_n(1,2,\dots,n)~ =~ \big\<\co_{1} \co_{2}\,\cdots\co_{n}\big\>\,.
\ee
 $SU(4)$ invariance  requires that the total number of (suppressed) indices
is a multiple of 4, i.e.   $\sum_{i=1}^n r_i\,=\,4m$.  Furthermore each index value $a=1,2,3,4$ must appear $m$ times among the operators $\co_{i}$.  The integer $m$ determines the N$^K$MHV class by $K = m - 2.$

In general one considers complex null momenta $p^\mu$, described by a bi-spinor $p^{\a\dot{\b}} = |p]^\a \<p|^{\dot{\b}}$.
For real momenta, when angle and square spinors are related by complex conjugation, each  $A_n(1,2,\dots,n) $ describes a physical amplitude in which particles in the final  state have positive null $p^\m$ and particles in the initial state have
negative null $p^\m$.  In scattering theory the $S$-matrix describes particle states in the  limit of infinite past and future in which wave packets separate and interactions can be neglected.   Therefore the SUSY charges that act on asymptotic states are determined by the free field limit of the transformation rules of the field theory.

In this section it is convenient to define chiral supercharges
 $Q^a \equiv - \e^\a\,Q^a_\a$ and
$\widetilde{Q}_a \equiv \e_{\dot{\a}} \widetilde{Q}^{\dot{\a}}_a$, which include contraction with the  anti-commuting parameters $\e^\a, \e_{\dot{\a}}$
of SUSY transformations.
The commutators of the operators $Q^a$
and $\widetilde{Q}_a$ with the various annihilators are given by:
\bea
\begin{array}{rcl}
\big[\tQ_a,B\big] &=& 0\, ,\\[2mm]
\big[\tQ_a,B^b\big] &=& \<\epsilon\, p\>\,\d^b_a\,B\, ,\\[2mm]
 \big[\tQ_a,B^{bc}\big]
&=&\<\epsilon\,p\> \, 2! \, \d^{[b}_a\,B^{\raisebox{0.6mm}{\scriptsize$c]$}}\, ,\\[2mm]
\big[\tQ_a,B^{bcd}\big] &=&\<\epsilon\, p\>\, 3!\, \d_a^{[b} B^{\raisebox{0.6mm}{\scriptsize$cd]$}}\, ,\\[2mm]
\big[\tQ_a,B^{bcde}\big] &=& \<\epsilon\,p\>\, 4!\, \d_a^{[b} B^{\raisebox{0.6mm}{\scriptsize$cde]$}} \, ,
\end{array}
\hspace{8mm}
\begin{array}{rcl}
[Q^a,B] &=& [p\,\epsilon]\,B^a\, , \\[2mm]
\big[Q^a,B^b\big] &=& [p\,\epsilon]\, B^{ab}\, , \\[2mm]
\big[Q^a,B^{bc}\big]
&=&[p\,\epsilon]\, B^{abc}\, , \\[2mm]
 \big[Q^a,B^{bcd}\big] &=& [p\,\epsilon]\,B^{abcd}\, , \\[2mm]
\big[Q^a,B^{bcde}\big] &=& 0\, ,
\end{array}
\lab{n4tQ}
\eea
Note that $\tQ_a$  raises the helicity of all operators and involves
the spinor angle bracket $\<\e\,p\>$.  Similarly, $Q^a$
lowers the helicity and spinor square brackets $[p\, \e]$ appear.

The Ward identities that relate S-matrix elements are obtained from
\bea
0 &=& \big\<\,[\,\tQ_a\,,\, \co_{1} \co_{2} \,\cdots  \co_{n}\,]\,\big\>
\, ,
\lab{tqward}\\[1ex]
0&=& \big\<\,[\,Q^a\,,\, \co_{1} \co_{2} \,\cdots  \co_{n}\,]\,\big\>\,.\label{qward}
\eea
The overall expressions vanish because supercharges annihilate
 the
 vacuum state.  One then obtains concrete relations among amplitudes by moving the supercharges to the right, using the appropriate entry from \reef{n4tQ} to evaluate $[\tQ_a, \co_{i}]$ or $[Q^a,\co_{i}]$.  To obtain a non-trivial relation, the product of operators $\co_{1} \,\cdots \co_{n}$ must contain an odd number of fermions.  There are further constraints from $SU(4)$ invariance. In \reef{tqward}.  the distinguished index value $a$ must appear $m+1$ times among the $\co_{i}$, while other index values appear $m$ times each.   Similarly, in \reef{qward},  the index value $a$ must appear $m-1$ times and other index values $m$ times each.  One thus sees that the Ward identities relate amplitudes within an N$^K$MHV class.

At the MHV level, the SUSY Ward identities give very simple and transparent relations. For example, consider
\begin{equation}
\begin{split}
 \lab{mhv1}
0 &~=~ \bigl\< \,[\,\tQ_1\,,\,--B^1 +\cdots+\,]\,\bigr\>\\[1ex]
&~=~ \<\e 1\> \bigl\< B^{234}-B^1+\cdots+\bigr\> \,\,+\, \<\e 2\> \bigl\<- B^{234}B^1 +\cdots+\bigr\> \,\,+\, \<\e 3\> \bigl\<--+ +\cdots +\bigr\>\,\,,
\end{split}
\end{equation}
where we used that negative helicity gluons `$-$' transform as $[\tQ_{1},B^{{\bf 1}234}_{i}]=\<\e i\>B^{234}_{i}$, while  positive helicity gluons `$+$' are annihilated by the supercharge,  $[\tQ_1,B_{i}]=0$\,.
There are three contributions on the right hand side
 of~(\ref{mhv1}):
the first two are gluon pair amplitudes and the last one is the $n$-gluon MHV amplitude.  However, there are two linearly independent choices of the SUSY spinor $\< \e|$.  If we choose $\<\e | \sim \<2 |$, then \reef{mhv1} yields the relation
\be \lab{mhv2}
 \bigl\<B^{234} -B^1 +\cdots +\bigr\> ~=~ \frac{\< 2 \,3\>}{\<1\, 2\>}\,
 \bigl\<--++\cdots +\bigr\>\,.
\ee
between a gluino pair amplitude and the $n$-gluon amplitude.
If we choose  $\<\e | \sim \<1 |$, then we find a similar relation for the other gluon pair amplitude.
For every set of operators in  $\<[\tQ_1, \co_{1}\,\cdots\co_{n}]\>$
in which the index $a=1$ appears three times and the
indices~$2,3,4$
 twice each, the Ward identity contains three terms.  By choice of $\<\e |$ one obtains two independent relations  similar to \reef{mhv2}.  By combining the various relations, one can show that any MHV $n$-point amplitude can be expressed as a rational function of angle brackets times the $n$-gluon amplitude
 $\<--++\cdots+\>\,$. Another fact about MHV amplitudes is that the $Q^a$ Ward identities are automatically satisfied when the relations from the $\tQ_a$ WI's are incorporated.
 These key properties of the MHV sector are best seen from the MHV generating function discussed in the next section.

The situation in the NMHV sector is very different, as we can see by examining  the Ward identity
\bea
0\!\!&=&\!\! \bigl\<\, [\,\tQ_1\,, \,- -- B^1++\,]\,\bigr\>\\[1ex]
\!\!&=&\!\! \<\e 1\> \bigl\<B^{234} \!-\!-B^1\!+\!+\bigr\> \,\,+ \,\<\e 2\> \bigl\<- B^{234}\!- \!B^1\!+\!+\bigr\>\,\,+\, \<\e 3\> \bigl\<-\!-\!B^{234} \!B^1\!+\!+\bigr\>\,\, + \, \<\e 4\> \bigl\<-\!-\!-\!+\!++\bigr\>  \,.\nonumber
\eea
Now there are four terms, while $\<\e |$ can take two independent values.
Thus,  one obtains two independent equations for four amplitudes, which is a linear system of
``defect two''.
Every
NMHV compatible
choice of the operators in $\<[\tQ_1, \co_{1}\,\cdots\co_{n}]\>$  produces a similar pair of linear equations.  Thus one obtains a large coupled set  of such relations, and the overall rank of the system is difficult to ascertain. This problem is
 indeed
best addressed in the language of superamplitudes, which we introduce in the next section.
Please read on.


\setcounter{equation}{0}
\section{Superamplitudes and their Symmetries }
\lab{secsusyR}

\subsection{Superamplitudes and supersymmetry constraints}
The annihilation operators of the 16 massless states --- gluons, gluinos, scalars --- of the $\cn=4$ supermultiplet can be
encoded in the `on-shell superfield'
\bea
  \label{Phi}
  \Phi = B + \eta_{a} \,B^a - \frac{1}{2!} \eta_{a}\,\eta_{b}\,B^{ab}
  - \frac{1}{3!} \eta_{a} \,\eta_{b} \,\eta_{c} \, B^{abc}
  +  \eta_{1} \,\eta_{2} \,\eta_{3} \,\eta_{4} \, B^{1234}\,,
\eea
in which the bookkeeping Grassmann variables $\eta_{a}$ are labeled by
$SU(4)$ indices $a,b,\ldots=1,2,3,4$.
The supercharges $\tilde{q}_{a} = \<\e p\> \eta_{a}$ and $q^a = [\e p] \tfrac{\partial}{\partial\eta_{a}}$
act on $\Phi$ by
multiplication or differentiation.  They `move'  operators  to the right or left in \reef{Phi} to reproduce the commutation relations \reef{n4tQ}.  The
 anticommutator of the two supercharges is $[\tilde{q}_a,q^b] = \d^b_a\<\e|p_i|\e]$ and thus realizes Poincar\'e SUSY.

The amplitudes for all $n$-point processes within a given N$^K$MHV class are collected into \emph{super\-amplitudes}
$\mathcal{A}_n(\Phi_{1},\dots,\Phi_{n})$, which are polynomials in the $\eta_{ia}$.
The superamplitudes we discuss here must be $SU(4)$ invariant.
 In particular, an   N$^K$MHV superamplitude is a degree $4(K\!+\!2)$ polynomial in the $\eta_{ia}$ in which each index value $a=1,2,3,4$ appears $(K\!+\!2)$ times in every monomial term.
Any desired amplitude can be  projected out from $\mathcal{A}_n$ by acting with the differential operators \cite{Bianchi:2008pu} that select the desired external state $B^{\dots}_{i}$ from each  $\Phi_{i}$. The total derivative order is $4(K\!+\!2)$.

The construction for $\mathcal{N}=8$ supergravity is completely analogous: the 256 massless states are encoded into superfields using Grassmann variables $\eta_a$ labelled by the global R-symmetry group $SU(8)$. The N$^K$MHV superamplitudes are degree $8(K\!+\!2)$ polynomials in the $\eta_{ia}$'s.  In the rest of this section we study the maximally supersymmetric
gauge and gravity theories ($\mathcal{N}=4$ and $\mathcal{N}=8$) jointly.

In \cite{Elvang:2009wd}  it is shown that the  SUSY Ward identities \reef{susyward} can be satisfied if superamplitudes are constructed from two basic ingredients. The first
 ingredient is the
 well-known Grassmann $\d$-function
\be \lab{deltq}
\d^{(2\cn)}(\tQd) ~\equiv~ \d^{(2\cn)}\Big(\sum_{i=1}^n |i\>\eta_{ia}\Big)
~ = ~ \frac{1}{2^\cn}\prod_{a=1}^{\cn}
\sum_{i,j}^n \<i\,j\>\, \h_{ia}\h_{ja}\,.
\ee
$\d^{(2\cn)}(\tQd)$ is fully supersymmetric.
 Indeed, it is clear that
$\widetilde Q_a\, \d^{(2\cn)}\big(\tQd\big)  =0$,
while momentum conservation ensures that $Q_a\, \d^{(2\cn)}\big(\tQd\big) =0$.
We will show below that $\d^{(2\cn)}(\tQd)$ is also $SU(\cn)$ invariant.

The $\d^{(2\cn)}$-function is the only element needed to construct
MHV superamplitudes. Note that it has the correct polynomial order, namely $2\cn $.  The $n$-point MHV superamplitude is simply given by
\be
  \lab{simple}
  \ca_n^\text{MHV}
  \,=\,
  \d^{(2\cn)}(\tQd)
  ~ \frac{ \<++\cdots+ - -\>   }{\<n-1,n\>^{\cn}} \, .
\ee
It has one `basis amplitude,' namely the pure gluon/graviton MHV amplitude $A_{n}(++\dots+--)$.
When the order-$2\cn$ differential
operator, which selects a given process, is
applied, $\cn$ angle brackets are produced from the
$\d^{(2\cn)}$-function, and the chosen amplitude is then ${\<..\>\cdots\<..\>}/\<n\!-\!1,n\>^{\cn}$ times the basis amplitude.

The second basic  ingredient that is needed to construct N$^K$MHV superamplitudes is the simple polynomial  $\pdan_{ijk,a}$ of \reef{pdanIntro}.
The  Schouten identity ensures $Q^a \,\pdan_{ijk,b} =0$, and this holds for \emph{any} choice of three lines $i,j,k$, adjacent or non-adjacent, independent of momentum conservation.

We write the N$^K$MHV superamplitude
\bea
 \label{Aform}
 \mathcal{A}_n^\text{N$^K$MHV}
  ~=~
  \delta^{(2\mathcal{N})}(\tQd)~P_{\mathcal{N}\times K}\,,
\eea
where $P_{\mathcal{N}\times K}$ is a polynomial of degree $\mathcal{N}\times K$ in the $\eta_{ia}$ variables. The delta-function \reef{deltq} ensures that
$\widetilde{Q}_a \mathcal{A}_n^\text{N$^K$MHV}= 0$. Since $Q^a$ commutes with the delta-function, the only remaining SUSY constraint is $Q^a P_{\mathcal{N}\times K}=0$. This is a non-trivial condition, but we  show that   its general solution  can be expressed in terms of products of the polynomials $m_{ijk,a}$. The  solution
depends on
the R-symmetry Ward identities, which we discuss next.

\subsection{R-symmetry}
\lab{s:Rsym}

To establish $SU(\mathcal{N})_R$ invariance of a function of the $\eta_{ia}$-variables it is sufficient to impose invariance under $SU(2)_R$ transformations acting on all choices of a pair of the $SU(\mathcal{N})_R$ indices $1,\dots, \cn$.
To be specific,  consider infinitesimal $SU(2)_R$ transformations in the $ab$-plane:
\be
  \lab{su2}
  \s_1:
  \Big\{
  \begin{array}{ccc}
  \delta_R \eta_{ia} &=& \theta \eta_{ib}\\
  \delta_R \eta_{ib} &=& \theta \eta_{ia}
  \end{array}
  \, ,
  \hspace{5mm}
  \s_2:
  \Big\{
  \begin{array}{ccc}
  \delta_R \eta_{ia} &=& -i \theta \eta_{ib}\\
  \delta_R \eta_{ib} &=& i \theta \eta_{ia}
  \end{array}
  \, ,
  \hspace{5mm}
  \s_3:
  \Big\{
  \begin{array}{ccc}
  \delta_R \eta_{ia} &=& \theta \eta_{ia}\\
  \delta_R \eta_{ib} &=& - \theta \eta_{ib }
  \end{array}
  \, .
\ee
Here $\theta$ is the infinitesimal transformation parameter.

As a warm-up to further applications, we show that the $\d^{(2\cn)}$-function \reef{deltq} is $SU(\cn)_R$ invariant; this implies that MHV superamplitudes necessarily preserve the full $R$-symmetry. Since any monomial of the form $\eta_{i1}\,\eta_{j2}\cdots \eta_{l\cn}$ is invariant under a $\sigma_3$-transformation, so is the $\d^{(2\cn)}$-function. A $\s_1$-transformation in the 12-plane gives
\be
  \lab{dd}
  \delta_R\, \big( \d^{(2\cn)}(\tQd) \big)
  ~=~
  \frac{\theta}{2^{\cn-1}}
  \Big(
   \sum_{i,j=1}^n \< ij\> \, \eta_{i1}\, {\bf\eta_{j2}}~
   \sum_{k,l=1}^n \< kl\> \,{\bf \eta_{k2} \, \eta_{ l2}} \Big)
   \Big(\prod_{a=3}^\mathcal{N}
  \sum_{k',l'=1}^n \< k'l'\> \eta_{k'a}\, \eta_{l'a}  \Big)~+ ~\dots
 ~~=~~ 0\, .~~~
\ee
Anticommutation of the (highlighted) Grassmann variables antisymmetrizes the sum over $j,k,l$ and $\< ij\>\< kl\>$ then vanishes by Schouten identity.
The ``$+\dots$" stands for independent terms from $\d_R$ acting   on $\eta_{k2}$ and $\eta_{l2}$. These terms can  be treated the same way. Invariance under  $\sigma_2$-transformations follows directly from $\sigma_{1,3}$ invariance and needs no further proof.

The R-symmetry  constraints play an important role in the analysis of the SUSY Ward identities beyond the MHV level.

The analysis of the R-symmetry Ward identities also leads to a set of new \emph{cyclic identities} for amplitudes. The identities encode relationships among amplitudes with the same types of external states, but with their R-symmetry indices distributed in different ways.  An example is the
 following 4-term relation among $\cn=4$ SYM NMHV amplitudes with gluinos $\l$ and scalars $s$:
\bea
  \nonumber
  0 &=&
  A_{6}( \l^{123} \l^{\bf 3} \l^{123} \l^{\bf 4} s^{1{\bf 4}} s^{2{\bf 4}} )
  ~+~ A_{6}( \l^{123} \l^{\bf 4} \l^{123} \l^{\bf 3} s^{1{\bf 4}} s^{2{\bf 4}} ) \\[1mm]
  &&
  ~+~A_{6}( \l^{123} \l^{\bf 4} \l^{123} \l^{\bf 4} s^{1{\bf 3}} s^{2{\bf 4}} )
  ~+~ A_{6}( \l^{123} \l^{\bf 4} \l^{123} \l^{\bf 4} s^{1{\bf 4}} s^{2{\bf 3}} )    \, .
  \lab{6ptexI}
\eea
We call this a cyclic identity because the four boldfaced $SU(4)$ indices are cyclically permuted.

\setcounter{equation}{0}
\section{Basis expansion of superamplitudes in $\cn=4$ SYM}
\label{secNMHV}

We outline the strategy used to solve the SUSY- and R-symmetry Ward identities and construct a particular basis for the amplitudes at  NMHV level for   $\mathcal{N}=4$ SYM.
We then present the  representations of superamplitudes using this basis. We emphasize results below and leave full details to App. \ref{app1}.

\subsection{Strategy for solving the SUSY Ward identities}\label{secstrategy}

The initial form of the $\cn=4$ NMHV superamplitude is
\be
 \label{n4start}
 \mathcal{A}_n^\text{NMHV}  ~=~  \delta^{(8)}(\tQd)~P_{4}\,.
\ee
Our task is to construct a minimal basis for all 4th order Grassmann polynomials $P_4$ that are $SU(4)$ invariant and satisfy $Q^a P_4=0$.   Let's get to work.

\begin{enumerate}
\item First consider the constraints of $SU(4)$ R-symmetry invariance discussed in
Sec.~\ref{s:Rsym}.
The $\sigma_3$-transformations require $P_4$ to be a linear combination of $\eta_{i1}\,\eta_{j2}\,\eta_{k3}\,\eta_{l4}$ monomials, so we write
\be
  \lab{Namp}
  P_4 ~= \sum_{i,j,k,l=1}^n q_{i j k l}\,
 \eta_{i 1} \, \eta_{j 2} \,  \eta_{k 3} \,  \eta_{l 4} \, .
\ee
The action of the $\sigma_1$-rotation
 in the $12$-plane gives
\begin{equation}
    \d_R\, (q_{ijkl}\, \eta_{i1}\,\eta_{j2}\,\eta_{k3}\,\eta_{l4}) = \theta \, q_{{\bf ij}kl}(\h_{\bf i2}\h_{\bf j2}+\h_{\bf i1}\h_{\bf j1})\eta_{k3}\,\eta_{l4}\,.
\end{equation}
This quantity must vanish; hence the coefficients $q$ must be symmetric in indices $i$ and $j$,
$q_{ijkl}=q_{jikl}$. A similar argument for any generator of $SU(4)_R$ implies
 that $q_{ijkl}$ is a totally symmetric tensor.

\item The superamplitude \reef{n4start} includes the $\delta^{(8)}$-function as a factor,  so the 8 conditions it imposes can be used  to eliminate a total of $8$ distinct $\eta_{ia}$, namely any choice of two $\eta_{ia}$ for each $a$.
A convenient choice (which we make) is to eliminate the 4+4 Grassmann variables associated with lines $n\!-\!1$ and $n$. Then~$P_{4}$ will then not depend on $\eta_{n-1,a}$ and $\eta_{na}$, and we write
\be
  \lab{step2}
 P_4 = \frac{1}{\<n\!-\!1,n\>^4} \sum_{i,j,k,l=1}^{n-2} c_{i j k l}\,
 \eta_{i 1} \, \eta_{j 2} \,  \eta_{k 3} \,  \eta_{l 4} \, .
\ee
The $c_{ijkl}$'s are linear combinations of the $q_{ijkl}$'s; we will not need their detailed relationship.
 The coefficient $1/\<n\!-\!1,n\>^4$ in~(\ref{step2}) could be absorbed by a redefinition of the $c_{i j k l}$, but we keep it for later convenience. As in step 1, R-symmetry requires the $c_{ijkl}$'s to be fully symmetric, so the number of needed inputs at this stage is $(n\!-\!2)(n\!-\!1)n(n\!+\!1)/4!$\,.

It is a consequence of our choice to eliminate $\eta_{n-1,a}$ and $\eta_{na}$ that  {\em all basis amplitudes have negative helicity gluons on lines $n\!-\!1$ and $n$.}

\item  The polynomial $P_4$ in \reef{step2} satisfies  the Ward identity  $Q^a P_4 = 0$  if and only if
the linear relations \mbox{$\sum_{i=1}^{n-2}\, [\eps i ] \, c_{ijkl}  = 0$} hold for any triple $jkl$.
Since $\eps^\a$ is a 2-component spinor, there are two independent constraints which  allow us to eliminate a choice of two lines $s$ and $t$ completely from the indices of the $c_{ijkl}$'s in $P_4$.
This is analogous to the use of the $\widetilde{Q}_a$ Ward identities to eliminate two sets of $\eta_{ia}$-variables in step 2, and a consequence is that \emph{lines $s$ and $t$  are positive helicity gluons in all basis amplitudes.}
In the following we choose $s=n\!-\!3$ and $t=n\!-\!2$.

We rewrite $P_4$ in terms of $c_{ijkl}$'s with $i,j,k,l \ne n\!-\!3,n\!-\!2$ and find  that this naturally leads to the appearance of the  polynomials $m_{ist,a}$, defined in \reef{pdanIntro}. The result  (see appendix for details) is the following form of the NMHV superamplitude:
\begin{equation}
   \label{Xs}
      \ca_n^\text{NMHV}~=\sum_{1\le i\le j\le k\le l \le n-4} c_{ijkl} \,X_{(ijkl)}\qquad\text{ with }\quad X_{(ijkl)} \equiv \sum_{\mathcal{P}(i,j,k,l)} X_{ijkl}\,,
\end{equation}
where the $X_{ijkl}$ are $\eta$-polynomials of degree 12 that are annihilated by both $Q^a$ and $\widetilde{Q}_a$:
\be\label{defXs}
    X_{ijkl} ~\equiv~ \d^{(8)}\big(\tQ_a\big)
  ~ \frac{m_{i, n\dash3,n\dash2;1}\,m_{j, n\dash3,n\dash2;2}\,m_{k, n\dash3,n\dash2;3}\,m_{l, n\dash3,n\dash2;4}}{[ n-3,n-2]^4\<n-1,n\>^4}   \,.
\ee
The sum over permutations $\mathcal{P}(i,j,k,l)$ in the definition of $X_{(ijkl)}$ is over all \emph{distinct} arrangements of fixed indices $i,j,k,l$. For instance, we have $X_{(1112)}=X_{1112}+X_{1121}+X_{1211}+X_{2111}$. Likewise,
$X_{(1122)}$ contains the 6 distinct permutation of
 its indices, and $X_{(1123)}$ has 12 terms.\footnote{The number of distinct permutations of a set with repeated entries is a multinomial coefficient \cite{stanley}.}

\item
 The coefficients $c_{ijkl}$ with $1\leq i,j,k,l\leq n\!-\!4$ parameterize the most general SUSY and $R$-symmetry invariant NMHV superamplitude.
The last step is to relate $c_{ijkl}$ to actual amplitudes which then become the basis amplitudes. By direct  application of the appropriate Grassmann derivatives,  we find that each $c_{ijkl}$ is identified as a single amplitude
\be \lab{ampl}
   c_{ijkl}  \,=\,
   A_n \big( \{i,j,k,l\}++ -- \big)
   \,\equiv\,
   \big\<\!  \cdots B_{i}^1 \!\cdots
     B_{j}^2 \!\cdots B_{k}^3\! \cdots B_{l}^4\! \cdots\,
   +_{n\dash3}+_{n\dash2}-_{n\dash1}-_{n}
   \big\> \,,
\ee
with $\!1\!\le\! i\! \le\! j\! \le\! k\! \le\! l \!\le\! n\!-\!4$. Let us clarify the notation: $A_n \big(  \{i,j,k,l\} ++--\big)$ means that line $i$ carries
$SU(4)_R$ index 1, line $j$ carries index 2 etc. If $i=j$, this means that
the line carries both indices 1 and 2, and the notation $B_{i}^1 B_{i}^2$ should then be understood as $B_{i}^{12}$. Furthermore the dots indicate positive-helicity gluons in the unspecified positions, specifically any state $\ne i,j,k,l,n\!-\!1,n$ is a positive helicity gluon. For example,
$A_{10} \big(  \{1,1,2,4\} ++--\big) = A_{10}(B^{12} B^3 + B^4 ++|++--)$. For clarity, we have used a `$|$' to separate the first $n\!-\!4$ states from the last four gluon states, which are the same for all basis amplitudes.
\end{enumerate}

Our final result for the manifestly SUSY and R-symmetric $\cn=4$ SYM NMHV superamplitude is
\be
   \lab{step4}
   \ca_n^\text{NMHV} ~=
     \sum_{1\le i\le j\le k\le l \le n-4}
      A_n \big(   \{i,j,k,l\}++ --\big)
     ~X_{(ijkl)}   \, .
\ee
One might say that we have used the SUSY generators  $Q^a$ and $\widetilde{Q}_a$  to `rotate' two states, $n\!-\!3$ and $n\!-\!2$, to be positive helicity gluons and two other states $n\!-\!1$ and $n$, to be negative helicity gluons.
Any NMHV amplitude can be obtained from \reef{step4}
by applying the 12th-order Grassmann derivative that corresponds to its external states. The amplitude will then be expressed as a linear combination of the \mbox{$(n\!-\!4)(n\!-\!3)(n\!-\!2)(n\!-\!1)/4!$}
 independent basis amplitudes $A_n \big(   \{i,j,k,l\}++ --\big)$. The collection of these amplitudes is what we define as the {\em algebraic basis}.

Let us consider examples of superamplitudes in the basis \reef{step4}. For $n=5$ we have to distribute the four $SU(4)$-indices on $n-4 = 1$ lines: there is only one choice, namely to put them all on line 1, which then must be a negative helicity gluon.  Thus the 5-point NMHV superamplitude is described in terms of a single basis element $A_5 \big(   \{1,1,1,1\}++ --\big) = \big\< -++--\big\>$; this is of course not surprising, since the
5-point
NMHV sector is equivalently described as anti-MHV. The superamplitude takes the form
${\ca}_5^\text{NMHV} =  \big\< -++--\big\> \,\, X_{1111}$.

Next, let us write the 6-point superamplitude in the basis \reef{step4}. The four $SU(4)$ indices should now be distributed in all inequivalent ways on lines $1$ and $2$. There are five ways to do this --- $1111$, $1112$, $1122$, $1222$ and $2222$ --- giving five basis amplitudes. The 6-point NMHV superamplitude can thus be written
\bea
  \nonumber
   {\ca}_6^\text{NMHV} \!\!&=&\!\!
   \big\< -+++--\big\> \,\, X_{1111}
   ~+~ \big\<  \lambda^{123} \lambda^{4}++ --\big\>\, X_{(1112)}
   ~+~\big\< s^{12}\, s^{34} \,++--\big\>\,
   X_{(1122)}
   \\[1mm]
   &&~~~~~~
   +~  \big\<  \lambda^{1} \lambda^{234} ++--\big\> \,X_{(1222)}
   ~+~ \big\< +-++--\big\> \, X_{2222}
    \, .
    \lab{nmhv6}
\eea
Here, we use a notation where $\lambda$ denotes a gluino ($B^a$ or $B^{abc}$)  with the indicated $SU(4)_R$ indices, and $s^{ab}$ denotes the scalar $B^{ab}$.

The amplitudes of the algebraic basis used in \reef{step4} are of the schematic form $\<\co_1\co_2\cdots \co_{n-4}++--\>$.
The states   $\co_i$  can be any particles of the theory, subject to the NMHV level constraint that each $SU(4)$ index $a=1,2,3,4$ appears once among the $\co_i$.
As in  any vector space,
 there many other ways to specify a basis.
 One can choose any other set with the same number of  amplitudes, provided that they are  linearly independent under the SUSY and  R-symmetry Ward identities.  To verify linear independence of a putative set of basis amplitudes one can project them from the superamplitude \reef{step4}  using the appropriate differential operators and then check that the matrix which relates the new set to the original basis has maximal rank. Due to algebraic complexity, this check is best done numerically using a computer-based implementation of the superamplitude.

At the $6$-point NMHV level, for example, any
choice of
$5$ linearly independent
$\cn=4$ SYM
amplitudes form a valid basis that completely determines the superamplitude.  We have verified that the split-helicity amplitude $A_{6}(+++---)$ together with $4$ of its cyclic permutations is a suitable basis of $6$-point NMHV amplitudes. Similarly, there are pure-gluonic algebraic basis for NMHV amplitudes with $n=7$ and $n=8$ external legs.
At $n=9$, however, the 84 distinct gluonic amplitudes span a 69-dimensional subspace of the  70-dimensional algebraic basis. For $n>9$ the dimension of the algebraic basis even exceeds the number of pure-gluon amplitudes, which immediately rules out the possibility of a purely gluonic basis.

\subsection{Functional bases and single-trace amplitudes}
\lab{secEx1}

The  representation \reef{step4} contains a sum over basis amplitudes which are \emph{algebraically} independent under the symmetries we have imposed.  However,
 we have not yet included possible \emph{functional} relations among amplitudes, that is relations which involve reordering of  particle momenta.   The cyclic and reflection symmetries of single trace color ordered amplitudes are examples of such relations.

For amplitudes in the single-trace sector, the cyclic permutations  are functionally dependent; they can be computed from cyclic momentum relabelings. Thus the all-gluon {\em algebraic basis}  of single-trace $6$-point NMHV amplitudes discussed above reduces to a {\em  functional basis containing the single amplitude  $\<+++---\>$.} (Note that functional relations among amplitudes do not invalidate their use in an algebraic basis.)

For $n>6$, the functional  basis in the single-trace sector cannot consist of a single amplitude. Indeed, dihedral symmetry relates $2n$ amplitudes, which, for $n>6$, is smaller than the number of algebraic basis amplitudes.
For example, for $n=7$ dihedral symmetry generates a set of at most 14 amplitudes from any one given amplitude, but ${n-1 \choose 4}=15$ amplitudes are needed needed to form an algebraic basis.
It is an open problem to find a simple expression of the superamplitude in terms of the minimal functional basis. However, it is possible to write down superamplitudes whose algebraic basis amplitudes are pairwise functionally related by dihedral symmetry. We refer the reader to~\cite{Elvang:2009wd} for details of this construction.

\subsection{Beyond NMHV: superamplitudes and Young tableaux}

The NMHV basis amplitudes  $A_n(\{i,j,k,l\}++--)$ of \reef{step4} are labeled by four integers in the range
$1\!\leq\! i\!\leq\! j\!\leq\! k\!\leq\! l\!\leq\! n\!-\!4$. These numbers are conveniently arranged in  the semi-standard tableaux
\raisebox{1pt}{
$\framebox[4mm][c]{\scriptsize $i$\phantom{j}\hspace{-1mm}}
  \framebox[4mm][c]{\scriptsize $j$\phantom{j}\hspace{-1mm}}
  \framebox[4mm][c]{\scriptsize $k$\phantom{j}\hspace{-1mm}}
  \framebox[4mm][c]{\scriptsize $l$\phantom{j}\hspace{-1mm}}$}
of the  Young diagram with one row and four columns.
It was shown in \cite{Elvang:2009wd} that semi-standard Young tableaux provide the general organizing principle for  N$^K$MHV superamplitudes.  These superamplitudes can be written in the schematic form
\be
  \lab{cartoon}
  \ca_{n}^\text{N$^K$MHV} ~=~
  \sum_I A_I \, Z_I \, ,
\ee
in which index $I$ enumerates the $SU(n-4)$ semi-standard tableaux of the rectangular  Young diagram $Y$ with $K$ rows and $\cn$ columns. The number of such semi-standard tableaux is the dimension $d_Y$ of the $SU(n-4)$ irrep corresponding to the Young diagram $Y$.  For each tableau there is a basis amplitude $A_I$ and a manifestly SUSY and $SU(4)_R$-invariant $\eta$-polynomial $Z_I$. To illustrate this structure, we discuss the  N$^2$MHV superamplitudes of $\cn=4$ SYM.

The basis amplitudes of the $n$-point N$^2$MHV superamplitude
are labeled by $SU(n-4)$ semi-standard Young tableaux with two rows and four  columns,
\begin{equation}
\begin{tabular}{|c|c|c|c|}
  \hline
  $\!\!i_1\!\!\!$ & $\!\!j_1\!\!\!$ & $\!\!k_1\!\!\!$ & $\!\!l_1\!\!$ \\
  \hline
  $\!\!i_2\!\!\!$ & $\!\!j_2\!\!\!$ & $\!\!k_2\!\!\!$ & $\!\!l_2\!\!$ \\
  \hline
\end{tabular}\,\,\,.
\end{equation}
Each row is non-decreasing ($i_A\!\leq\! j_A\! \leq\! k_A\!\leq\! l_A$) and each column is strictly increasing ($i_1\!<\!i_2$\,, etc.). Each tableau corresponds to a basis amplitude
$A_n\bigl(\,
\bigl\{
  {}^{i_1 j_1  k_1  l_1}_{i_2 j_2  k_2  l_2}
\bigr\}
\,++--\,\bigr)$
with the specified gluons on the last four lines and with $SU(4)_R$ index $1$ on lines $i_1$ and $i_2$, $SU(4)_R$ index $2$ on lines $j_1$ and $j_2$, etc.   For example,
\begin{equation}
  \label{exN2}
  \begin{array}{l}
  \framebox[3.5mm][c]{\scriptsize 1}
  \framebox[3.5mm][c]{\scriptsize 1}
  \framebox[3.5mm][c]{\scriptsize 1}
  \framebox[3.5mm][c]{\scriptsize 3}\\[-2pt]
  \framebox[3.5mm][c]{\scriptsize 2}
  \framebox[3.5mm][c]{\scriptsize 2}
  \framebox[3.5mm][c]{\scriptsize 2}
  \framebox[3.5mm][c]{\scriptsize 5}
  \end{array}.
  ~~~~\longleftrightarrow~~~~
  A_9\bigl(
     \bigl\{
    {}^{1113}_{2225}
    \bigr\}
     ++--\bigr)
   ~=~
   A_9( \lambda^{123} \lambda^{123} \lambda^{4} + \lambda^{4}
  ++--) \, .
 \end{equation}
{From the hook rule \cite{hook} it follows that the}
\be\label{N2MHVcount}
  \text{\#(N$^2$MHV $n$-pt basis amplitudes)} ~=~
  {{\rm dim}_{SU(n-4)}} \hspace{-1mm}
  \begin{array}{l}
  \framebox[3.5mm][c]{\scriptsize \phantom{1}}
  \framebox[3.5mm][c]{\scriptsize \phantom{1}}
  \framebox[3.5mm][c]{\scriptsize \phantom{1}}
  \framebox[3.5mm][c]{\scriptsize \phantom{1}}\\[-2pt]
  \framebox[3.5mm][c]{\scriptsize \phantom{1}}
  \framebox[3.5mm][c]{\scriptsize \phantom{1}}
  \framebox[3.5mm][c]{\scriptsize \phantom{1}}
  \framebox[3.5mm][c]{\scriptsize \phantom{1}}
  \end{array}
  ~=~
  \tfrac{(n-5)(n-4)^2(n-3)^2(n-2)^2(n-1)}{4! \, 5!} \, .
\ee

The N$^2$MHV superamplitude can be written in terms of  basis amplitudes as
\begin{equation}
   \ca^{\text{N$^2$MHV}}_n ~=~
   \frac{1}{16}
     \sum_{{}^\text{semi-standard}_\text{~~tableaux Y}}\!\!\! (-)^Y
     A_n\bigl(\,
     \bigl\{
    {}^{i_1 j_1  k_1  l_1}_{i_2 j_2  k_2  l_2}
    \bigr\}
     ++--\bigr)~
      Z^{i_1 j_1 k_1 l_1}_{i_ 2j_ 2k_2 l_2}
     \,,
\end{equation}
where the $Z$'s are
 manifestly SUSY- and R-symmetry invariant $\eta$-polynomials similar to the
 $X$'s in~(\ref{defXs}), but contain eight instead of four powers of $m_{ijk,a}$. The $Z$-polynomials and the sign factor $(-)^Y$ are defined
 in~\cite{Elvang:2009wd}.

Let us comment on the detailed information contained in the semi-standard tableaux labels of the basis amplitudes. In the example \reef{exN2}, line labels $1,2,3,4,5$ appeared $3, 3, 1, 0, 1$ times, respectively. This is a particular (ordered) partition of the $8\!=\!3\!+\!3\!+\!1\!+\!0\!+\!1$ boxes of the Young diagram; each semi-standard tableau corresponding to a 3+3+1+0+1 partition of 8 corresponds to a process with the same particles types for the external states: states 1 and 2 are negative helicity gluinos, states 3 and 5 are positive helicity gluinos, and state 4 is a positive helicity gluon. How many independent basis amplitudes are there corresponding to this partition? --- In other words, how many $SU(4)$-inequivalent ways are there to arrange the two sets of $SU(4)$-indices on the two negative helicity gluinos and the two positive helicity gluinos? The answer to
this question is the combinatorial quantity called the \emph{Kostka number}.\footnote{The Kostka number $C_{\l;Y}$ depends on a Young tableaux $Y$ with $M$ boxes and a partition $\lambda$ of $M$. The partition $\l$ is a weight that dictates the number of times each number is used in the construction of the semi-standard tableaux of $Y$. The Kostka number  $C_{\l;Y}$ counts the number of semi-standard tableaux of $Y$ with weight $\l$.} For the partition $3\!+\!3\!+\!1\!+\!0\!+\!1$ of the 2-by-4 rectangular Young diagram, the Kostka number is 2: in addition to \reef{exN2} there is a second basis amplitude  with the same particle types on each external line,
namely
\begin{equation}
  \label{exN2b}
  \begin{array}{l}
  \framebox[3.5mm][c]{\scriptsize 1}
  \framebox[3.5mm][c]{\scriptsize 1}
  \framebox[3.5mm][c]{\scriptsize 1}
  \framebox[3.5mm][c]{\scriptsize 2}\\[-2pt]
  \framebox[3.5mm][c]{\scriptsize 2}
  \framebox[3.5mm][c]{\scriptsize 2}
  \framebox[3.5mm][c]{\scriptsize 3}
  \framebox[3.5mm][c]{\scriptsize 5}
  \end{array}
  ~~~~\longleftrightarrow~~~~
  A_9\bigl(
     \bigl\{
    {}^{1112}_{2235}
    \bigr\}
     ++--\bigr)
   ~=~
   A_9( \lambda^{123} \lambda^{124} \lambda^{3} + \lambda^{4}
  ++--) \, .
 \end{equation}
Note that a different ordering of the partition is also possible, namely
$8\!=\!3\!+\!3\!+\!1\!+\!1\!+\!0$. The Kostka number is independent of the ordering, so there are also two semi-standard tableaux associated with this second ordering; they are just obtained from those in \reef{exN2} and \reef{exN2b} by
exchanging $5$ and $4$.
The corresponding basis amplitudes have
a positive helicity gluino on line 4 and a positive helicity gluon on line 5.  The structure outlined in this example generalizes to characterize all basis amplitudes of N$^K$MHV superamplitudes. Further details can be found in \cite{Elvang:2009wd}.

\setcounter{equation}{0}
\section{Basis expansion of superamplitudes in $\cn=8$ supergravity}
\label{n8samps}

The generalization of the above results to $\mathcal{N}=8$ supergravity is straightforward. The MHV sector is particularly simple because the superamplitude contains only one basis amplitude which we take to be the $n$-graviton amplitude $M_n(--+\dots+)$. The superamplitude is the 16th order Grassmann polynomial
\be \lab{mhvsamp}
\cac^{\rm MHV}_n =
\d^{(16)}\Bigl(\sum_i | i\> \eta_{ai}\Bigr) \frac{M_n(--+\dots+)}{\<12\>^8}\,.
\ee
The
 amplitude  $M_n(--+\dots+)$ must be bose symmetric under exchange of helicity spinors for the two negative helicity particles and for any pair of positive helicity particles.  However the superamplitude must have full $S_n$ permutation symmetry, and so must the ratio $M_n(\ldots)/\<12\>^8$\,.

 At the N$^K$MHV level, the amplitudes of the algebraic basis  are now characterized by the $SU(n-4)$ semi-standard tableaux of a rectangular $8$-by-$K$ Young diagram. The SUSY- and R-invariant Grassmann polynomials $Z_I$ multiplying each basis amplitude are order $8K$; they are constructed as in $\mathcal{N}=4$, but with twice as many $\eta_{ia}$'s.  N$^K$MHV $n$-point superamplitudes must also have $S_n$ permutation symmetry
We now discuss the NMHV sector in more detail.

\noindent {\bf NMHV amplitudes in $\cn=8$ supergravity}\\
The identification of an algebraic basis
 in supergravity proceeds as in gauge theory and leads
to a representation of  NMHV
superamplitudes analogous to
\reef{step4}
namely
\be \lab{repgrav}
 \cm_n^\text{NMHV}\,=\!\!\!
  \sum_{1\le i \le j \le \dots \le v \le n\dash4}\!\!\!
  c_{ijklpquv}~
  X_{(ijklpquv)},
\ee
with symmetrized versions of the $Q^a$- and $\tQ_a$-invariant polynomial
\be \lab{xpoly8}
 X_{ijklpquv} ~=~ \d^{(16)}(\tQ_a)~
 \frac{\,\pdan_{i,n\dash 3,n\dash 2;1}\,\,\pdan_{j,n\dash 3,n\dash2;2}\,\cdots\,\pdan_{v,n\dash 3,n\dash 2;8}\,}{[n-3,n- 2]^8\<n-1,n\>^8 }\,\,.
\ee
 As in $\cn=4$ SYM, we can identify each
 coefficient $c_{ijklpquv}$ with an amplitude:
 \begin{equation}
    c_{ijklpquv}  ~=~
   M_n \big( \{i,j,k,l,p,q,u,v\}++ -- \big)
   ~\equiv~
   \big\<  \cdots B_i^1 \cdots
     B_j^2 \cdots\cdots  B_v^8 \cdots
   ++--
   \big\>  \, .
 \end{equation}
The notation $\{i,j,k,l,p,q,u,v\}$  indicates that line
 $i$ carries $SU(8)_R$ index 1, while line $j$ carries $SU(8)_R$ index 2, etc.
 If indices are identical, say $i=j$,
 the line in question carries both $SU(8)_R$ indices $1$ and $2$.

In gravity, as opposed to gauge theory, there is no ordering of the external states. Therefore amplitudes with the same external particles
 and the same $SU(8)_R$ charges are all related by momentum relabeling. For example,
\be
  c_{22222222}~=~
\<+_1 -_2 +_3 +_4 -_5 -_6\>  = \<-_2 +_1 +_3 +_4 -_5 -_6\>
  = \big(c_{11111111} ~~\text{with}~~p_1 \lra p_2\big)\,.
\ee
Since there are a total of eight $SU(8)_R$ indices $1,2,\dots,8$  distributed on these $n-4$ states, the number of functionally independent amplitudes cannot exceed the number of partitions of 8 into $n-4$ non-negative integers.

For example, for $n=6$ we have the partitions $[8,0]$\,, $[7,1]$\,, $[6,2]$\,, $[5,3]$ and $[4,4]$ corresponding to a reduced set of 5 functional basis amplitudes in the functional basis.  The 6-point superamplitude is then
\bea
  \nonumber
   {\cal M}_6^\text{NMHV} &=&~
   \Big\{~~
   \big\<    -\, + \,+ + - -\big\>\,\, X_{\,11111111\,}
   \,~+~ \big\<   \psi^{-}\psi^{+}  + + - -\big\>\,
   X_{(11111112)}
   \\[1mm]  \nonumber
   &&~\,
   +\big\<  v^-v^++ + - -\big\>\,
   X_{(11111122)}
   ~+~\big\< \chi^{-}\chi^{+} + + - -\big\>\,
   X_{(11111222)} \\[1mm]
   &&~\,
   +\frac{1}{2}\big\< \, \phi^{1234} \phi^{5678} ++  - -\big\>\,
   X_{(11112222)} \Big\}
   ~~+~~ \text{(1 $\leftrightarrow$ 2)}\, .
   \lab{SGNMHV}
\eea
Particle types are indicated by $\psi^+ = B^{1}$, $\psi^- = B^{2345678}$ etc, in hopefully self-explanatory notation. The ``$\,+~(1 \lra 2)$'' exchanges momentum labels 1 and 2 in the $X$-polynomials as well as in the basis amplitudes. The exchange does not introduce new basis functions, it only relabels momenta in the basis amplitudes written explicitly in \reef{SGNMHV}.

For the $n$-point NMHV superamplitudes,  the number of amplitudes in the functional basis is the number of partitions of the number 8 into $n-4$ bins:

\begin{tabular}{rcccccccccccccccc}
  $n~~= $ & 5 & 6 & 7 & 8 & 9 & 10 & 11 & $\ge$ 12 \\[1mm]
  $\text{basis count}~
  \,= $  & 1 & 5 & 10 & 15 & 18 & 20 & 21 ~& ~~~~22.
\label{count}
\end{tabular}

\vspace{1mm}
\noindent
The entry in the second line is the number of $n$-point amplitudes one needs to compute in order to fully determine the $n$-point NMHV superamplitude.  The saturation at $n=12$ occurs because the longest partition of $n-4=8$ is reached, namely the partition $[1,1,1,1,1,1,1,1]$. This partition corresponds to a basis amplitude with 8 gravitinos, two positive-helicity gravitons and two negative-helicity gravitons.
For $n>12$, one only adds further positive-helicity gluons to each partition. This does not change the count of
 functional basis amplitudes.

\vspace{2mm}
\noindent {\bf Minimal functional basis} \\
 In the functional basis discussed above, we have considered the functional dependence between algebraic basis amplitudes of the form
 $\<\co_1\cdots \co_{n-4}++--\>$. As in $\cn=4$ SYM, one can use a computer-based implementation of the superamplitude to check linear independence of a different set of algebraic basis amplitudes that are not all of this form. Although the superamplitude then generically takes a very complicated form in terms of these basis amplitudes, this approach is still convenient as the amplitudes in the new algebraic basis can exhibit a larger functional dependence. At the $6$-point NMHV level, for example, we verified that the $9$ algebraic basis amplitudes $\<\co\co++--\>$ can be replaced by an algebraic basis that consists of the pure-graviton amplitude  $\<---+++\>$, together with 8 inequivalent permutations of its external gravitons. Therefore, {\em at the $6$-point NMHV level, the minimal functional basis consists of a single amplitude, $\<---+++\>$.}
For $n>6$, however, the dimension of the algebraic basis exceeds the number of pure-graviton amplitudes, which rules out the possibility of a pure-graviton functional basis.

\vspace{2mm}
\noindent {\bf Examples} \\
Let us illustrate the solution to the Ward identities in a few explicit examples.  Consider first the amplitude with 2 sets of 3 identical gravitinos, $\<\psi^{1234567}\,  \psi^8  \psi^8  \psi^8  \psi^{1234567}  \psi^{1234567}\>$. Applying the corresponding Grassmann derivatives \cite{Bianchi:2008pu} to the superamplitude \reef{SGNMHV}, we find
\bea
 \nonumber
 &&\hspace{-2cm}
 \<\psi^{1234567}\, \psi^8 \, \psi^8 \, \psi^8\,  \psi^{1234567} \psi^{1234567}\>\\[1mm]
 &&=
 \frac{1}{[34]\<56\>} \,
 \Big\{
  ~ \< 2|3+4|1 ] ~\big\<    -\, + \,+ + - -\big\>
   + s_{234}  ~\big\<   \psi^{-}\psi^{+}  + + - -\big\>
 \Big\} \, ,
 \lab{ex1}
\eea
where $s_{234} = - (p_2 +p_3+p_4)^2$.
This particular $\cn=8$ amplitude agrees with the 6-gravitino amplitude
$\big\< \psi^{-}\psi^{+} \psi^{+} \psi^{+} \psi^{-} \psi^{-}\big\>$ in the
truncation of the $\cn=8$ theory to $\cn = 1$ supergravity. In fact the relation \reef{ex1} is a special case of the ``old'' solution to the $\cn=1$ SUSY Ward identities \cite{Grisaru:1977px,Bianchi:2008pu}.

An example which does not reduce to $\cn=1$ supergravity is obtained by interchanging
 the $SU(8)_R$ indices 7 and 8 on states 1 and 2 in the 6-gravitino amplitude. The result is another 6-gravitino amplitude whose expression in terms of basis amplitudes is found to be
\bea
 \nonumber
 &&\hspace{-2cm}
 \<\psi^{1234568}\, \psi^7 \, \psi^8 \, \psi^8\,  \psi^{1234567} \psi^{1234567}\>\\[1mm]
 &&=
 -\frac{1}{[34]\<56\>} \,
 \Big\{
  ~  s_{134}  ~\big\<   \psi^{-}\psi^{+}  + + - -\big\>
   +  \< 1|3+4|2 ] ~\big\<    v^-\, v^+ \,+ + - -\big\>
 \Big\} \, .
 \lab{ex2}
\eea
This example could be interpreted as the solution to the SUSY Ward identities in $\cn=2$ supergravity.

Our final example contains two distinct scalars and four gravitinos:
\bea
 \lab{ex3}
 &&\hspace{-1cm}
 \<\phi^{1238}\, \phi^{4568} \, \psi^7 \, \psi^8\,  \psi^{1234567} \psi^{1234567}\>\\[1mm]
 &&=
 \frac{\< 2|1+4|3 ]}{[34]^2\<56\>} \,
  \Big\{
  ~   [14]    ~\big\<   \chi^{-}\chi^{+}  + + - -\big\>
     + [24]   ~\big\<   \phi^{1234}\phi^{5678}  + + - -\big\>
 \Big\}  ~-~(1 \lra 2) \, .
 \nonumber
\eea
We have checked the solutions \reef{ex1}, \reef{ex2}, and \reef{ex3} numerically at tree level using the MHV vertex expansion, which is valid
\cite{Bianchi:2008pu}
for the specific $\cn=8$ amplitudes considered here. Of course the relations \reef{ex1}, \reef{ex2}, and \reef{ex3} hold in general, at arbitrary loop order.

\setcounter{equation}{0}
\section{Application: superamplitude approach to counterterms}

A theoretical development without application is like a bicycle without wheels.  For this reason we now review the application \cite{Elvang:2010jv} of the basis expansions of superamplitudes to study candidate counterterms for $\cn=8$ supergravity.  The $\cn =8$ theory \cite{deWit:1977fk,Cremmer:1979up} is the maximal supergravity theory in $D=4$ spacetime dimensions, and
the idea was expressed quite early that it might have favorable ultraviolet
properties.
Recent support for this idea has come from the remarkable calculations of \cite{BernetalFiniteness} based on the generalized unitarity method~\cite{Unitarity,Bern:2011qt,Carrasco:2011hw},
 which showed that 4-point amplitudes are  UV finite in 3-loop and 4-loop order.  It is interesting to ask
whether this situation continues to higher number of external legs and higher loop order. If not, then we ask in which amplitudes  and at which loop order might the first divergence occur?

In four dimensions,
the coupling constant $\kappa$ of perturbative quantum gravity theories has dimensions of length.  Dimensional analysis then shows that the degree of divergence increases with the loop order $L$,
 but is independent of the number of external legs $n$.
A  logarithmic divergence at loop order $L$ would require a
 local counterterm of dimension $\D =  2(L+1)$. Here, we define the dimension of an operator in a slightly non-standard manner as its ``power-counting dimension'', \ie as the number of derivatives plus $\half$ the number fermionic fields that it contains.\footnote{External line factors carry dimension $1/2$ for fermions,  but are dimensionless for bosons.}    The counterterms of $n$-point graviton amplitudes must respect general coordinate invariance and thus take the form $\int d^4x \sqrt{-g} D^{2k}R^n$. This is a schematic form in which the index contractions and distribution of covariant derivatives
on the curvature tensor are not specified.  A counterterm of this form
 has dimension $2(k+n)$, and could describe a UV divergence at loop level $L = n+k-1$.

In $\cn=8$  supergravity,  the
lower spin fields are unified with gravity, so counterterms must contain a supersymmetric completion involving those fields, which we denote very schematically by
$\int d^4x \sqrt{-g}( D^{2k}R^n + \dots)$.
Terms in the linearized SUSY completion contribute to $n$-graviton and other $n$-particle processes, while non-linear terms contribute to various processes with more than $n$ external particles. Little is known about the component form of the supersymmetrization of these operators, nor is it needed in the approach \cite{Elvang:2010jv}  we now review.

The approach of  \cite{Elvang:2010jv} focuses on   matrix elements of candidate counterterm operators. If an operator $D^{2k}R^n$ has at least a linearized  supersymmetric
completion then the $n$-particle matrix elements it generates must obey the SUSY Ward identities discussed in Secs.~\ref{s:susywis}.
 Furthermore, and crucially,
the leading $n$-point matrix elements of any counterterm must be local;  this means that they must not have any poles in their dependence on momenta $p_i$; gauge invariance then implies that they are polynomials
in the spinor brackets $\<i\,j\>,~ [k\,l]$.  The matrix elements must also be $SU(8)$ invariant.

In many cases,  the requirements of locality, $SU(8)$ symmetry and SUSY are incompatible.  This proves that no supersymmetrization exists, and the operator cannot occur in the perturbation series of $\cn =8$ supergravity.

In other cases,  the  Ward identities and locality are compatible.  The operator is then linearly $\cn=8$ supersymmetrizable and $SU(8)$ symmetric. It is accepted as a potential candidate counter\-term pending a study of the questions of nonlinear SUSY and  the low energy theorems of the $E_{7(7)}$ symmetry.
  Low energy theorems were considered in \cite{Elvang:2010kc,Beisert:2010jx}
  (see also~\cite{Brodel:2009hu}), which we review in Sec. \ref{secE77}. For operators compatible with locality, linearized
  SUSY and R-symmetry, our method constructs their general matrix elements explicitly. In particular, this allows us to determine the multiplicity\footnote{
  This multiplicity is the number of independent linearized supersymmetrizations
  of an operator, including distinct index contractions. Operators are considered dependent if they are related by the linearized equations of motion; in that case their leading matrix elements are identical.}
of the operator.

Two simple features facilitate the matrix element approach.
\begin{itemize}
\item   There are constraints on matrix elements from dimensional analysis and particle helicities,  and these become particularly powerful for local matrix elements. Spinor brackets $\<i\,j\>,~[k\,l]$ have mass
dimension 1, and all terms in a possible  supersymmetrization  $D^{2k}R^n +\ldots$ have dimension $2(k+n)$.   Thus   any matrix element of this operator must be a sum of monomials which each contain  $2(k+n)$ angle and square brackets.  The particle helicity constraint arises from the little group scaling property \cite{Witten:2003nn}
\bea\label{lgs}
  m_n(\dots, t_i |i\> , t_i^{-1} |i], \dots) = t_i^{-2h_i}\, m_n(\dots, |i\> , |i], \dots )\,,
\eea
which holds for each particle $i$.
This determines the difference $a_i-s_i = -2h_i$ between the number of angle spinor $|i\>$ and square spinor $|i]$ factors in each term of $m_n$.

\item
The index contractions  of an operator $D^{2k}R^n+ \dots $ can be organized according to the N$^K$MHV classification of its $n$-point matrix elements. This is
 possible because on-shell the Riemann tensor $R_{\mu\nu\rho\sigma}$  splits
into a totally symmetric 4th rank spinor $R_{\a\b\g\d}$ and its conjugate $\bar{R}_{\dot{\a}\dot{\b}\dot{\g}\dot{\d}}$, which communicate to gravitons of opposite helicity. Terms in
$D^{2k}R^n$ with 2 factors of $R$ and $(n-2)$ factors of $\bar{R}$ contribute to the MHV graviton matrix element while $R^3\,\bar{R}^{n-3}$ is the NMHV part and so on.  This separation persists in the SUSY completion, because the SUSY Ward identities relate amplitudes within  a given N$^K$MHV sector.
\end{itemize}

The import of this is that we can use the basis expansions of N$^K$MHV superamplitudes discussed in Sec.~\ref{secNMHV}. The first step is to determine the  basis amplitudes: each basis amplitude is  constructed as the most general polynomial in angle and square brackets  consistent with
helicity-scaling, Bose-symmetry, and dimensional requirements.
Two polynomials are identical if they are related by momentum conservation and Schouten identities. A systematic way to construct a complete set of polynomials subject to these requirements is to consider the polynomials as elements of a quotient ring $\mathbb{P}\big[ \<ij\>, [ij]\big]/\mathbb{I}$, where $\mathbb{P}\big[ \<ij\>, [ij]\big]$ is the ring of all polynomials in angle- and square brackets and $\mathbb{I}$ is the ideal generated by the polynomial conditions for momentum conservation and Schouten identities. Gr\"obner basis techniques can then be used to find a basis for the vector space of all polynomials, of fixed degree $\Delta$ and given
 little-group scaling weights~(\ref{lgs}), in the quotient ring. Linear combination of these basis elements then constitute the most general expression for a given basis amplitude.

The above construction ensures that the basis amplitudes are local.
The next step is to
demand
that all amplitudes produced by the corresponding superamplitude are also local; this requires that the factors in the denominator of  (\ref{mhvsamp}) and
(\ref{repgrav})-(\ref{xpoly8})
cancel in all amplitudes. As we shall see, this is a nontrivial constraint.

If the poles in non-basis amplitudes do not cancel for any admissible choice of basis polynomials,  then the operator under study does not have an acceptable supersymmetrization and is ruled out. If, on the other hand, the method determines one or more sets of basis amplitudes that do lead to a  local and permutation symmetric superamplitude, then each set  yields  an independent linear supersymmetrization of the operator. In practice, it is difficult to explicitly verify locality of all non-basis amplitudes. However, it was shown in~\cite{Elvang:2010jv} that {\em any superamplitude with local basis matrix elements and full permutation symmetry produces local matrix elements for any process.} This reduces the difficult process of checking locality of non-basis amplitudes to the much more practical check of permutation symmetry of the full superamplitude.

The matrix element method cannot predict whether an accepted candidate counterterm corresponds to an actual divergence in the perturbative S-matrix of $\cn=8$ supergravity.
At loop levels $L= 3,4$,
the results of  \cite{BernetalFiniteness} show evidence for cancelations beyond those associated with $\cn=8$ SUSY and this situation may persist.
 As we discuss in
 Sec.~\ref{secE77}, the additional constraints from $E_{7(7)}$ explain and predict  the absence of any UV divergences below $7$-loop order.

\subsection{Candidate MHV counterterms}  \label{secMHV}

\paragraph{Ruling out $R^n$ for $n \ge 5$:}
To see how the method of \cite{Elvang:2010jv} works, let us ask whether the operator $R^n$ has a linearized supersymmetrization
 at the MHV level.
Its $n$-point matrix element
 $m_n(--+\dots+)$
must be a polynomial  with the spinor powers   $|1\>^4,\, |2\>^4$ and $|i]^4,\, i=3,\dots,n$\,, which are the minimal powers consistent with the helicity weights  $-2h_i = 4,4,-4,-4,\dots$\,.  With these minimal powers the total dimension $2n$ is saturated, so the basis matrix element in \reef{mhvsamp} must take the form
\be \lab{Rnmhv}
m_n(--+\dots+) ~=~  \<12\>^4 f_n(|3],|4],\ldots |n])\,.
\ee
The function $f_n$ is an order $2n-4$ polynomial in square brackets, and
depends only on square spinors $|i]$ for positive helicity gravitons, \ie $i \ge 3$. It  must also be bose symmetric, but we will not need this information. The MHV superamplitude is obtained by inserting this matrix element in \reef{mhvsamp}, which then reads
\be \label{samprn}
\cac^{\rm MHV}_n =
\d^{(16)}\Bigl(\sum_i | i\> \eta_{ai}\Bigr)\frac{f_n(|3],|4],\ldots |n])}{ \<12\>^4}\,.
\ee

The basis matrix element
 $m_n(--+\dots+)$
is local, but we must test whether all other matrix elements obtained by differentiation of
\reef{mhvsamp} are also local.   We will examine the $n$-graviton matrix element with the negative helicity gravitons on lines 3 and 4.  To `project out'  a negative helicity graviton on line $i$, one applies the 8th order Grassmann derivative
 \be \lab{gravderiv}
 \Pi_i \equiv \prod_{a=1}^8 \frac{\pa}{\pa \h_{ia}}\,.
 \ee
We thus find the permuted matrix element
\be \lab{Rn34}
m_n(++ -- + \dots +) =
\Pi_3\Pi_4\cac^{\rm MHV}_n =
\frac{\<34\>^8}{\<12\>^4}\times f_n(|3],|4],\ldots |n])\,.
\ee
We now show that the non-locality
in $\<12\>$ does not cancel for $n\geq 5$.
To do this we introduce a complex variable $z$ and
evaluate \reef{Rn34} for the shifted spinors
\be  \lab{shift}
| i \>  \to |\hat{i}\> ~=~ | i \> + z c_i |\xi\>\,, \qquad\qquad  i =1,2,5\,,\qquad \sum_i c_i |i] =0\,,
\ee
with all other angle spinors and all square spinors unshifted.\footnote{An  ``anti-holomorphic'' shift similar to
 \reef{shift}  has been used  in \cite{Risager:2005vk,Elvang:2008na,Elvang:2008vz,Cohen:2010mi}  to prove the CSW recursion relations \cite{Cachazo:2004kj}
 (see~\cite{Brandhuber:2011ke} for a review in this issue).}
The quantity $|\xi\>$ is an arbitrary reference spinor. The shift affects only the denominator  in \reef{Rn34}, so the right-hand side has an uncanceled 4th order pole in $z$.  Therefore the amplitude  $m_n(++ -- + \dots +)$ is non-local, even with the input of the most general basis polynomial in \reef{Rnmhv} which satisfies the scaling constraints.   $R^n$  MHV counterterms for $n\ge 5$ are therefore ruled out!

The condition $\sum_i c_i |i]=0$ in \reef{shift} is needed so that the shifted spinors satisfy momentum conservation.  There are non-vanishing choices of the constants $c_i$ only if at least 3 lines are shifted.  Therefore the shift
does not work when there are only 4 external particles,
 and cannot be used to rule out $R^4$. We will discuss the
$R^4$ counterterm shortly.

 \paragraph{Ruling out $D^2 R^n$, $D^4R^n$, and $D^6R^n$ for $n\ge 5$:}
Next consider potential MHV counterterms $D^{2k}R^n$. Their overall dimension is $2(k+n)$.   To satisfy the scaling constraints we must construct basis polynomials with the minimal powers used in \reef{Rnmhv}
plus $2k$ additional matched pairs $|i\>,~|i]$ for any choice of up to $2k$ lines.  The basis amplitude thus contains $4+k$ angle brackets. When shifted, it becomes a  polynomial in $z$ of order no greater than $4+k$.  For $k<4$, this is not enough to cancel the 8th order pole from the factor
$1/\<12\>^8$
in the superamplitude.  The shift argument thus rules out the MHV counterterms $D^2 R^n,~ D^4R^n,~D^6R^n$ for $n \ge 5$.

\paragraph{Ruling in $D^8R^n$}
The analysis has ruled out MHV operators $D^{2k}R^n$ for $k<4$.
It may not be immediately clear whether the bound
$k<4$ is a limitation of  method  or  fact.  We now settle that question in favor of fact by exhibiting that the bound is saturated: we do this by
explicit construction of
an  MHV superamplitude for the counterterm $D^8R^n$.
The 8 angle brackets required by  scaling weights allow the factor $\<12\>^8$ which directly cancels the singular factors in
\reef{simple},
leaving the
manifestly local superamplitude~\cite{Elvang:2010jv}
\be \lab{d8rnmhv}
\cac^{\rm MHV}_n =
\d^{(16)}\Bigl(\sum_i | i\> \eta_{ai}\Bigr) \Big[c_1 \big([12]^2 [23]^2 \cdots [n1]^2 + \text{perms}\big)
 +c_2 \big(([12] [34] \cdots [n-1,n])^4 + \text{perms}\big)\Bigr] \,.
\ee
The second term only exists if $n$ is even, but the first is valid for all $n$.
For $n=4$ these two terms are linearly dependent through the Schouten identity.
For $n=6$ the two terms are independent, and there are no other independent contributions.\footnote{Other structures may become available when $n$ is sufficiently large.}

\paragraph{$D^{2k}R^4$ are allowed for all $k\ne 1$:} \label{d2kr4}
Consider a possible $R^4$ counterterm.
There is only one local expression for its basis amplitude with the correct dimension and weights, and it is
\be
 \label{m4R4}
  m_4(1^-,2^-,3^+,4^+)= \<12\>^4[34]^4\, .
\ee
This form also appears, for example,  in \cite{Kallosh:2008mq}. The better known \cite{Gross:1986iv}  form  $m_4(1^-,2^-,3^+,4^+)= s\, t\, u\, M_4^{{\rm tree}}$ is equivalent to  \reef{m4R4} using momentum conservation
\be \lab{momcons}
\<y x\>[x z] ~=~ -\!\!\sum_{i \ne x} \<y i\>[i z]\,.
\ee
The resulting superamplitude is
\be
\cac_4^{\rm MHV} = \d^{(16)}\Bigl(\sum_i | i\> \eta_{ai}\Bigr)\frac{[3\,4]^4}{\<1\,2\>^4} \,.
\ee
Using \reef{momcons} one can show that \emph{all} matrix elements obtained from it are local. Indeed, this means that $\cac_4^{\rm MHV}$  is local.

For all $k \ge 0$, the allowed polynomial form of the basis matrix element of $D^{2k}R^4$ is  $m_4(-\,-\,+\,+) = g_{D^{2k}\!R^4}\<1\,2\>^4[3\,4]^4 $,  so the  superamplitude is
\be
\cac_4^{\rm MHV} = \d^{(16)}\Bigl(\sum_i | i\> \eta_{ai}\Bigr)g_{D^{2k}\!R^4}(s,t,u)\frac{[3\,4]^4}{\<1\,2\>^4}\,.
\ee
$g$ is an order $k$ symmetric polynomial in $s,t,u$, for example
$g_{R^4} =1$, $g_{D^{2}\!R^4} = s+ t+ u=0$, $g_{D^{4}\!R^4} = s^2+t^2+u^2$,  $g_{D^{6}\!R^4} = s^3+t^3+u^3$ [\citen{GVR},\citen{Elvang:2010jv}]. Since  $g_{D^{2}\!R^4}=0$, the 4-loop counterterm $D^2R^4$ is ruled out. For all other $k$, $D^{2k}R^4$ is allowed by $\cn=8$ SUSY and $SU(8)$-symmetry.

\subsection{Candidate NMHV counterterms}
\lab{secNMHVb}

The extension of the matrix element method to the NMHV level is based on the  superamplitudes \reef{repgrav},  which are Grassmann polynomials of order 24.
 For each basis amplitude, we input the most general  polynomial in spinor brackets  consistent with
 helicity-scaling, Bose-symmetry, and dimensional requirements.
  The superamplitudes guarantee that
individual matrix elements, obtained
by Grassmann differentiation, are related by the appropriate SUSY Ward
identities.   Since the Ward identities are under control, we can proceed to study   to test if the non-basis matrix elements produced from the superamplitudes
are also local.

In  \cite{Elvang:2010jv},  NHMV level $R^n$ and $D^2R^n$  counterterms were ruled out by a shift argument similar to  that used at the MHV level in
 Sec.~\ref{secMHV}.
  Supersymmetric
NMHV  counterterms $D^4R^n$,
on the other hand, can be constructed.
The NMHV bound is weaker than in  the MHV sector where independent $D^{4}R^n$ and $D^{6}R^n$ counterterms were also ruled out.  In this review we discuss only the case $n=6$.

\subsubsection{No $R^6$ and $D^2R^6$ NMHV counterterms}

\label{secR6D2R6}
The 6-point superamplitude was discussed in Sec.~\ref{n8samps}.  It is convenient here to work with a more schematic form of \reef{SGNMHV}.
There are 9 terms in the symmetrized sum, so we write
\be \lab{n86pt}
{\cal C}^{\rm NMHV}_6 =  \sum_{j=0}^8 m^{(j)}X_ {(j)}\,.
\ee
The $m^{(j)}$ indicate the basis amplitudes of \reef{SGNMHV} in sequential order, e.g.  $  m^{(0)}  = m_6(-+++--),~m^{(1)}=m_6(\psi^-\psi^+++--), \dots , ~m^{(8)} =m_6( +-++--)$. Note that $m^{(5)},\dots,m^{(8)}$ are
in the $1\lra 2$ exchanged part of \reef{SGNMHV}.   The  $X_{(j)}$ are the corresponding symmetrized 24th order Grassmann polynomials.  They are the symmetrizations of the polynomials in \reef{xpoly8},  but with $n=6$.  They contain the singular factor $1/([34]\<56\>)^8$ which will be important shortly.

The basis matrix elements  of the  superamplitude  describing a
possible supersymmetrization of the operator $D^{2k}R^6$
 must be local expressions of mass dimension $2(k+6)$, so the total number of angle and square spinors is $\sum_i(a_i +s_i) = 4(k+6)$.   Helicity weights determine the  difference  $\sum_i(a_i -s_i)= -2\sum_i h_i =  0$  for any basis element of \reef{SGNMHV}.  Thus each basis matrix element is a product of $\sum_i a_i = 6+k$ angle and $\sum_i s_i= 6+k$ square brackets.

Using a suitable complex shift, we now show that
when $k=0,1$
the potential pole factor $1/ \<56\>^8$ \emph{cannot cancel}
in  the permuted $6$-graviton matrix element $m_6(--+++-)$  obtained from the superamplitude~(\ref{SGNMHV}). We project out $m_6(--+++-)$ from the superamplitude by applying the Grassmann derivatives defined in \reef{gravderiv} for negative helicity graviton lines, obtaining
\begin{equation}
    m_6(--+++-)~=~\,
    \Pi_1\Pi_2\Pi_6
    {\cal C}_6^{\rm NMHV}
  \lab{perm6}
~=~ \frac{1}{\<56\>^8} \sum_{j=0}^8 {8\choose j} \,\<26\>^{8-j}\<16\>^j\, m^{(j)}\,.
\ee
The eight angle brackets in the numerator come from derivatives
of the Grassmann $\delta^{(16)}$ in the $X$-polynomials~(\ref{xpoly8}). The factor $1/[34]^8$   in \reef{SGNMHV} cancels in \reef{perm6}  because differentiation
of the $m_{ijk,a}$ polynomials produces compensating factors in all terms.
The binomial coefficients appear because of the symmetrization of labels in the $X$-polynomials.

Consider now the effect  of the
holomorphic $3$-line shift of angle spinors as in~(\ref{shift}), but acting on the spinors $|3\>$, $|4\>$, and $|5\>$.  Spinor brackets $\<q \,q'\>$ are invariant under this shift unless they involve at least one spinor from the set $|3\>$, $|4\>$, $|5\>$. Shifted brackets are linear in $z$.  The denominator of \reef{perm6} has an 8th order pole in $z$,  but the brackets $\<26\>$ and
 $\<16\>$  in the numerator do not shift. The only potential $z$ dependence in the numerator comes
from the $6+k$ spinor brackets in the basis matrix elements $m^{(j)}$.
The poles cannot cancel in any linear combination of basis elements if  they contain fewer than 8 shifted angle brackets.  Thus the counterterm is ruled out if $6+k<8$; hence for $ k= 0,\,1$\,.

\subsection{$7$-loops: Explicit NMHV superamplitudes for  $D^4 R^6 $ }\label{secexplicit}

In
Sec.~\ref{secR6D2R6},
we used a shift argument to rule out NMHV counterterms $D^{2k}R^n$ with $k\leq 1$.
The result $k \le 1$ is  an actual bound for NMHV operators, not just a limitation of the method. Indeed,  two independent supersymmetric $7$-loop NMHV operators $D^4 R^6$ were constructed in \cite{Elvang:2010jv,Beisert:2010jx}, and it was also shown that precisely two such operators exist.

In fact it is quite simple to write down a new type of representation of the two superamplitudes.   Since their matrix elements contain products of 8 angle and 8 square brackets,  one can conjecture that they can be written in the form
\be \label{newsuper6}
{\cal C}_6 = \delta^{(16)}(\tQd)\delta^{(16)}(Q) P_{24}(\eta_{ia})\,,
\ee
where $P_{24}(\eta_{ia})$ is a 24th order $SU(8)$ invariant polynomial in the
$\eta$'s and  $\delta^{(16)}(Q)$ is the Grassmann differential operator
\begin{equation}
        \delta^{(16)}(Q)=\prod_{a=1}^8\sum_{i<j}\,[ij]\, \frac{\partial^2}{\partial \eta_{ia}\partial \eta_{ja}}\,.
\end{equation}

 It is not hard to write down two candidate polynomials
for the explicit superamplitudes. Their independence was verified numerically.
For example, one can choose
\begin{equation}
\begin{split}\label{D4R6super}
    {\cal C}_{D^4\!R^6}&=\delta^{(16)}(\tQd)\delta^{(16)}(Q)\bigl[(\varphi_1,\varphi_2)(\varphi_3,\varphi_4)\,(\varphi_5,\varphi_6)+\text{perms}\bigr]\,,\\
    {\cal C}_{D^4\!R^6{}'}&=\delta^{(16)}(\tQd)\delta^{(16)}(Q)\bigl[(\varphi_1,\,\varphi_2,\,\varphi_3,\,\varphi_4,\,\varphi_5,\,\varphi_6)+\text{perms}\bigr]\,.
\end{split}
\end{equation}
The sums in~(\ref{D4R6super}) run over all inequivalent permutations of the external state labels $i$ of the $\varphi_i$, and the $\varphi$-products are defined as
\begin{equation}
\begin{split}
    (\varphi_i,\varphi_j)\equiv\,\,&\epsilon^{a_1a_2a_3 a_4 b_1b_2b_3 b_4}\prod_{t\!=1}^4\eta_{ia_t}\eta_{jb_t}\,,\\[-.2ex]
    (\varphi_i,\varphi_j,\varphi_k,\varphi_l,\varphi_m,\varphi_n)\equiv\,\,&
    \epsilon^{a_1a_2b_1b_2b_3 b_4c_1c_2}\epsilon^{c_3c_4d_1d_2d_3 d_4e_1e_2}
    \epsilon^{e_3e_4f_1f_2f_3 f_4a_3a_4}
    \!\prod_{t\!=1}^4
    \eta_{ia_t}\eta_{jb_t}\eta_{kc_t}\eta_{ld_t}\eta_{me_t}\eta_{nf_t}\,.
\end{split}
\end{equation}
Of course, the choice of contractions is not unique,
and it is only through the  independently established multiplicity count in \cite{Beisert:2010jx} that we know the two contractions given in~(\ref{D4R6super}) to be sufficient.

\subsection{Summary: Potential counterterms}
\label{summary}
We have excluded MHV and NMHV operators $D^{2k}R^n,~n>4,$ with $k<4$ and $k<2$, respectively. Since divergences in $L$-loop amplitudes correspond to counterterms of dimension $2L+2$, this translates to the bounds
\bea
  \label{bound1}
  \text{no MHV:}~~~~L < n+3 \, , ~~~~~~~~~~~
  \text{no NMHV:}~~~~L < n+1 \, ,  ~~~~~~~~~
  (n>4)
\eea
for the \emph{exclusion} of $\cn=8$ SUSY and $SU(8)$-invariant operators $D^{2k}R^n$ in each class.  The explicit construction of the set of MHV superamplitudes~(\ref{d8rnmhv}) for  $D^{8}R^n$  and of the 7-loop NMHV superamplitudes~(\ref{D4R6super}) for  $D^4R^6$ show that the bounds are optimal.

The bounds \reef{bound1} also apply \cite{Elvang:2010jv} to the existence of \emph{non-gravitational} counterterms such as  $D^{2k}\phi^{m}+\ldots$ whose supersymmetrizations do not include any purely gravitational terms. Furthermore, it was conjectured in \cite{Elvang:2010jv} and proven in \cite{Drummond:2010fp}, that the bound
\bea
  \label{bound2}
  \text{no N$^K$MHV:}~~~~L < n+3 - 2K \, ,   ~~~~~~~~~
  (n>4)
\eea
holds for \emph{all}  N$^K$MHV operators of dimension $2L+2$. Operators  below this bound are not compatible with $\cn=8$ SUSY and $SU(8)$ R-symmetry. Charts of available operators and their multiplicities were given in \cite{Elvang:2010jv,Beisert:2010jx},  and a concise chart that summarizes the available counterterms is given in table~\ref{tabchart}.

\begin{table*}[t!]
\includegraphics[width=15.5cm]{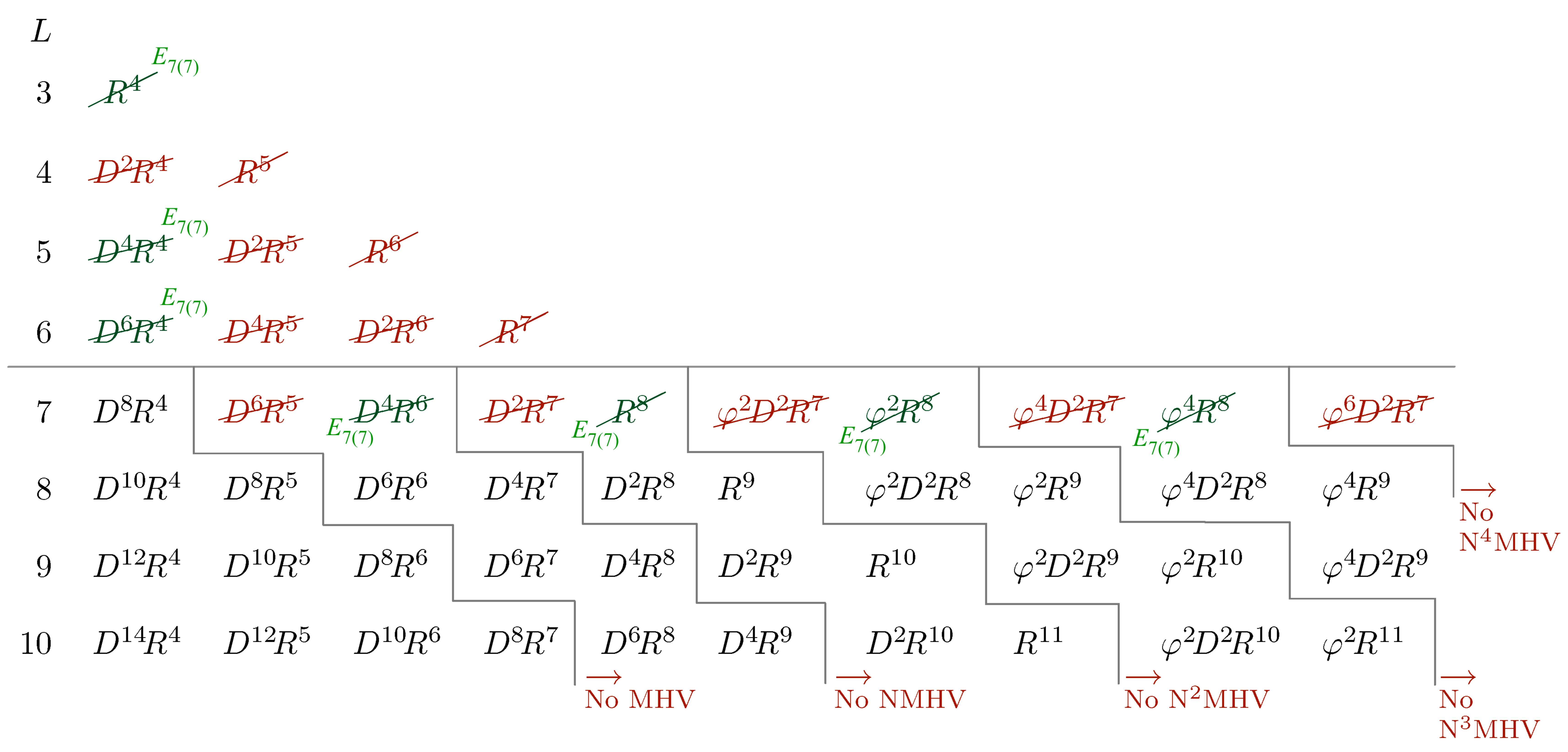}
\caption{
Potential counterterms in $\cn=8$ supergravity.
The crossed-out operators are excluded. At loop order $L<7$, only three operators are allowed by SUSY and R-symmetry, namely $R^4$, $D^4R^4$ and $D^6R^4$. However, these are not compatible with nonlinear $E_{7(7)}$ symmetry and this means that there are no available counterterm operators below $7$-loop order.
For $L \ge 7$, both the matrix element method of this review and
an analysis of the representations of the superalgebra $SU(2,2|4)$ were
used.  The second method is described in \cite{Beisert:2010jx}, and a more detailed version of the chart appears there.
}
\label{tabchart}
\end{table*}

The 4-point operators $R^4$, $D^4R^4$ and $D^6R^4$, with the rather simple 4-point superamplitudes discussed in Sec.~\ref{d2kr4}, are the only operators below the $7$-loop level that are consistent with SUSY and $SU(8)_R$.
However, we show in the next section that these operators are  ruled out as possible counter\-terms
by the nonlinear
 $E_{7(7)}$ symmetry.  Thus the combined SUSY, R- and $E_{7(7)}$-symmetries leave no candidate counterterms for $L<7$.

\subsection{$E_{7(7)}$ constraints on counterterms}\label{secE77}
Since the operators $R^4$, $D^4R^4$, and $D^6R^4$ are compatible with SUSY and  $SU(8)_R$ symmetry, more information is needed to rule out these operators as potential counterterms of $\cn=8$ supersymmetry.
We show now,
following the analysis of~\cite{Elvang:2010jv} and its extension in  \cite{Beisert:2010jx}, that $R^4$, $D^4R^4$, and $D^6R^4$ are incompatible with continuous $E_{7(7)}$ symmetry. Recall that $E_{7(7)}$ symmetry is spontaneously broken to its maximally compact subgroup $SU(8)_R$; the 70 scalars in the spectrum of $\cn=8$ supergravity are the Goldstone bosons associated with this symmetry breaking. It has been argued \cite{Bossard:2010dq} that the $E_{7(7)}$ symmetry is also a symmetry at loop level, and from this
it follows from  the ``soft-pion theorem'' that the matrix elements of an admissible counterterm must vanish when the momentum of any external scalar is taken to zero.\footnote{
Non-vanishing SSL's  from external  line insertions occur in pion physics, but not with the cubic vertices of $\cn=8$ \cite{Bianchi:2008pu,ArkaniHamed:2008gz,Kallosh:2008rr}.}
 Our matrix-element approach to counterterms is thus ideally suited to address the question of $E_{7(7)}$-compatibility.

The leading $4$-point matrix elements of $R^4$, $D^4R^4$, and $D^6R^4$ take the form
\begin{equation}
    \big\<\co_1\co_2\co_3\co_4 \big\>_{D^{2k}R^4}~=~g(s,t,u) \,\,\big\<\co_1\co_2\co_3\co_4\big\>_{\rm SG}\,,
\end{equation}
where $\<\cdot\cdots\>_{\rm SG}$ is the tree-level  $\cn =8$ supergravity amplitude with the same choice of external states, and the function $g(s,t,u)$ is given by
\begin{equation}\label{gs}
\begin{split}
    g_{R^4}&=1\,,\qquad\quad~\, g_{D^4R^4}=~s^2+t^2+u^2\,,\qquad\quad g_{D^6R^4}=s^3+t^3+u^3\,.
\end{split}
\end{equation}
Tree-level supergravity amplitudes have vanishing  single-soft scalar limits (SSL's), so the SSL's of the $4$-point matrix elements also vanish for all $4$-point operators:
\begin{equation}
    \lim_{p_\varphi\to 0} \big\<\varphi\,\cdots\big\>_{D^{2k}R^4} ~=~ \lim_{p_\varphi\to 0}  g(s,t,u) \,\big\<\varphi\,\cdots\big\>_{\rm SG}~=~0\,.
\end{equation}
Thus  we need
to consider higher-point matrix elements of $D^{2k}R^4$ to rule out these operators.

Specifically, we study the soft scalar limit of the 6-point NMHV matrix elements  $\<\hp\hp\hm\hm\varphi\bar\varphi\>_{D^{2k}\!R^4}$. The external states are two pairs of opposite helicity gravitons and two conjugate scalars.
These matrix elements contain local terms from $n$th order field monomials in the nonlinear SUSY completion of $D^{2k}R^4$ as well as
non-local
pole diagrams in which one or more lines of the operator are off-shell and communicate to tree vertices from the classical Lagrangian. It is practically impossible to calculate these matrix elements with either Feynman rules (because the non-linear supersymmetrizations of $D^{2k}\!R^4$ are unknown) or recursion relations (because no valid ones are known). Instead we use the $\alpha'$-expansion of the closed string tree amplitude to obtain the desired matrix elements.

At tree level, the closed string effective action takes the  form
\begin{eqnarray}
\label{effS}
     S_{\rm eff}&=&S_\text{SG}-2\alpha'^3\zeta(3)e^{\dash 6\phi}R^4-\zeta(5)\alpha'^5e^{\dash 10\phi}D^4R^4+\tfrac{2}{3}\alpha'^6\zeta(3)^2e^{\dash12\phi}D^6R^4+\dots\,.
\end{eqnarray}
Couplings of the dilaton $\phi$ break the $SU(8)$ symmetry of the supergravity theory to $SU(4)\!\times\!SU(4)$ when $\alpha'>0$,
so matrix elements constructed from $S_{\rm eff}$ do not directly correspond to the  desired $SU(8)$-invariant operators. As explained in \cite{Elvang:2010kc}, an {\em $SU(8)$-averaging procedure} can be used to extract the $SU(8)$ singlet contribution from the string matrix elements. Specifically, the $SU(8)$ average of the $\<\hp\hp\hm\hm\varphi\,\bar\varphi\>_{e^{\dash(2k+6)\phi}D^{2k}\!R^4}$ matrix elements from string theory is
\begin{eqnarray}
 \label{su8avg}
 \<\hp\hp\hm\hm\varphi\bar\varphi\>_{\rm avg} = \tfrac{1}{35} \<\hp\hp\hm\hm\varphi^{1234}\varphi^{5678}\>
 - \tfrac{16}{35} \<\hp\hp\hm\hm\varphi^{123|5}\varphi^{4|678}\>
 + \tfrac{18}{35} \<\hp\hp\hm\hm\varphi^{12|56}\varphi^{34|78}\>\,.
\end{eqnarray}
The 3 terms on the right side correspond to the 3 inequivalent ways to construct scalars from particles of the $\cn =4$ gauge theory, namely from gluons, gluinos, and $\cn=4$ scalars. There are 35 distinct embeddings of $SU(4)\!\times\!SU(4)$ in $SU(8)$. Averaging is sufficient to give the matrix elements of the $\cn =8$ field theory operator $R^4$ and $D^4R^4$. For $D^6R^4$, a further correction is necessary and is discussed below in
 Sec.~\ref{s:D6R4}.

\subsubsection{From open strings to closed strings}
All closed string tree amplitudes in the work \cite{Elvang:2010kc,Beisert:2010jx}
are obtained via KLT~\cite{Kawai:1985xq} from the open string tree amplitudes of \cite{StSt}.
The KLT relations express closed string amplitudes $M_n$ as products of  `left' and `right' sector open string amplitudes $A_n^{(L)}$ and $A_n^{(R)}$; schematically
\bea
  \lab{KLT}
  M_n = \sum_\mathcal{P} f(s_{ij}) \,A_n^{(L)}\,A_n^{(R)}\,.
\eea
The sum is over permutations of different orderings of the external states of the open string amplitudes. The functions $f(s_{ij})$ involve a product of $n\!-\!3$ factors $\sin(\alpha' \pi s_{ij})$, where $s_{ij}$ are Mandelstam variables.
The decomposition of $\cn=8$ states into products of two $\cn=4$ states (L- and R-movers) is described in detail in \cite{Bianchi:2008pu}. To obtain the $6$-point closed string amplitude $\<\hp\hp\hm\hm\varphi\bar\varphi\>$ for the three independent pairs of conjugate scalars $\varphi$, $\bar \varphi$\,, the open-string amplitudes presented in~\cite{StSt} are not sufficient. Instead, SUSY Ward identities are needed to express the desired open string amplitudes in terms of the ones given\footnote{In~\cite{Brodel:2009hu}, Ward identities were also used for this purpose.} in~\cite{StSt};  but this is precisely the problem that we solved in
Sec.~\ref{secNMHV}!
We can use the superamplitude~(\ref{nmhv6}) to express the basis amplitudes
in terms of the open-string tree amplitudes amplitudes of~\cite{StSt}. Then we project out any other desired process from the superamplitude~(\ref{nmhv6}). Via KLT, we can then obtain any $6$-point NMHV closed-string tree amplitude.
The resulting closed string amplitudes confirm the structure and coefficients of (\ref{effS}).


\subsubsection{$3$- and 5-loop counterterms $R^4$ and $D^4R^4$}
At order $\alpha'^3$ and $\alpha'^5$, the $SU(8)$-average (\ref{su8avg}) of the string theory
amplitudes directly  gives the matrix elements of the unique $SU(8)$-invariant supersymmetrization of $R^4$ and $D^4R^4$, respectively.  The result is a complicated
 non-local expression, but its single-soft scalar limit is very simple and local, viz.
\begin{equation}\label{soft}
   \lim_{p_6\to 0}\<\hp\hp\hm\hm\varphi\bar\varphi\>_{R^4}=-\tfrac{6}{5}[12]^4\<34\>^4\,, \qquad
    \lim_{p_6\to 0}\<\hp\hp\hm\hm\varphi\bar\varphi\>_{D^4R^4}
    =-\tfrac{6}{7}[12]^4\<34\>^4\sum_{i<j}s_{ij}^2\,.
\end{equation}
Since these SSL's are non-vanishing, the operators $R^4$ and $D^4R^4$ are incompatible with continuous $E_{7(7)}$ symmetry.

\subsubsection{6-loop counterterm $D^6R^4$}
\label{s:D6R4}
The single-soft scalar limit of the $SU(8)$-singlet part of the closed string matrix element at order $\alpha'^6$, obtained by the $SU(8)$-averaging~(\ref{su8avg}), is
\begin{equation}\label{softstringyD6R4}
    \lim_{p_6\to 0}\<\hp\hp\hm\hm\varphi\bar\varphi\>_{(e^{\dash12\phi}D^6\!R^4)_{\rm avg}}=-\tfrac{33}{35}[12]^4\<34\>^4\sum_{i<j}s_{ij}^3\,.
\end{equation}
At order $\alpha'^6$, it is important to realize that the
 $6$-point NMHV  closed string amplitudes receive contributions not only
from
 $e^{-12\phi}D^6R^4$, but also from pole diagrams with two 4-point vertices of $e^{-6\phi}R^4$. Since no dimension 8 operator ($R^4$ nor $e^{-6\phi}R^4$)  is present in $\cn=8$ supergravity, its contributions must be removed from the string tree matrix element in order to extract the matrix elements of the supergravity operator $D^6R^4$.  The removal process must be supersymmetric.

We first compute the $R^4\!-\!R^4$ pole contributions to the 6-graviton NMHV matrix element   $\<\hm\hm\hm\hp\hp\hp\>$ as follows.  This amplitude has dimension 14.  Factorization at the pole determines the simple form
\begin{equation}\label{dandidit}
 \<12\>^4[45]^4\<3|P_{126}|6]^4/P_{126}^2 + 8 ~{\rm permutations}\,,
\end{equation}
up to a local polynomial.
The 9 terms correspond to the 9 distinct
non-vanishing
 3-particle pole diagrams.  The result~(\ref{dandidit}) is then checked by computation of the Feynman diagrams from the $R^4$ vertex \cite{grosssloan}.
 As the non-linear supersymmetrization of $R^4$ may contribute additional local terms, we also consider adding
the most general  gauge-invariant and bose-symmetric  polynomial of dimension $14$ that can contribute to $\<\hm\hm\hm\hp\hp\hp\>$, namely
\begin{equation}  \label{hediditagain}
\big(\<12\>\<23\>\<31\>[45][56][64]\big)^2 P_{123}^2\,.
\end{equation}

To incorporate SUSY, we
recall from the discussion of the ``minimal functional basis'' in
 Sec.~\ref{n8samps},
 that there is an algebraic basis
consisting of  $\<\hm\hm\hm\hp\hp\hp\>$ and 8 distinct permutations of the states.  In this basis we write a superamplitude ansatz as the sum of the pole amplitude (\ref{dandidit}) plus a multiple of (\ref{hediditagain}).   We then impose full $S_6$ permutation symmetry on the ansatz.  This fixes the coefficient of the polynomial~(\ref{hediditagain}) to vanish and   determines the SUSY completion of the desired pole diagram uniquely.

Finally we project out the scalar-graviton matrix element from this superamplitude and subtract its (properly normalized) soft scalar limit from
~(\ref{softstringyD6R4}) to obtain
\begin{equation}\label{mkdidit}
    \lim_{p_6\to 0}\<\hp\hp\hm\hm\varphi\bar\varphi\>_{D^6R^4}
    =-    \tfrac{60}{35}
    [12]^4\<34\>^4\sum_{i<j}s_{ij}^3\,.
\end{equation}
This is
 the single-soft scalar limit of the unique independent $D^6R^4$ operator in $\cn=8$ supergravity. Since the limit does not vanish, the operator $D^6R^4$ is
 incompatible with continuous $E_{7(7)}$ symmetry.

As explained in
 Sec.~\ref{summary},
 $R^4$, $D^4R^4$ and $D^6R^4$ are the only local supersymmetric and $SU(8)$-symmetric operators for loop levels $L\leq 6$ \cite{Drummond:2003ex,Elvang:2010jv,Drummond:2010fp}. Hence $\cn=8$ supergravity has no potential counterterms that satisfy the continuous $E_{7(7)}$ symmetry for $L\leq 6$.  We stress that  string theory is used as a tool to extract $SU(8)$-invariant
 matrix elements that must agree with the
 matrix elements of the $\cn=8$ supergravity operators $R^4$, $D^4R^4$ and $D^6R^4$ because each of these operators is unique.
 No remnant of string-specific dynamics remains in the final results.

It is instructive to check whether the scalar matrix elements of the NMHV
operators $D^4R^6$ have non-vanishing single-soft scalar limits.  Scalar matrix elements can be projected
 from the two superamplitudes~(\ref{D4R6super}), and direct computation shows that SSL's do not vanish for any linear combination of the two. Thus
these potential 7-loop, $n=6$ NMHV counterterms  are also  incompatible with the continuous $E_{7(7)}$ symmetry. The same conclusion holds for all other independent higher-point $7$-loop operators $R^8$, $\varphi^2R^8$, $\varphi^4R^8$, $\ldots$\,, as was shown in~\cite{Beisert:2010jx} using a different, complementary method (see also~\cite{Kallosh:2010kk}).

\subsubsection{Matching of soft scalar limits to automorphism analysis}

The non-vanishing of the single-soft scalar limits of the matrix elements for $R^4$, $D^4R^4$ and $D^6R^4$
 show
that these operators are not compatible with $E_{7(7)}$ symmetry. This conclusion was also suggested in a very different approach by Green, Miller, Russo and Vanhove \cite{Eisenstein}, who studied the moduli-dependence in $D$-dimensions of the above three supersymmetric operators. They found that the moduli-dependent function $f(\varphi)$ that appears in the non-linear completion $f(\varphi)D^{2k}R^4$  of these operators has to obey a certain Laplacian eigenvalue equation. The results for the single-soft scalar limits
 were used in \cite{Elvang:2010kc,Beisert:2010jx} to compute
the relative coefficient of the constant and the quadratic terms of the functions $f(\varphi)$,
and it
matched exactly the
 prediction of \cite{Eisenstein}
for all three cases $R^4$, $D^4R^4$ and $D^6R^4$.

\setcounter{equation}{0}
\section{Superamplitudes without maximal $R$-symmetry}
\label{s:lessRsym}
In this section we consider $\cn=8$ SUSY superamplitudes that transform non-trivially under the $SU(8)_R$ symmetry. Our techniques apply to  superamplitudes that preserve any subgroup of $SU(8)$, or even break $SU(8)$ completely, but we focus on superamplitudes that are invariant under an $SU(4)\!\times\! SU(4)$ subgroup.
These superamplitudes are relevant for two reasons:
\begin{itemize}
\item
As discussed in
Sec.~\ref{secE77},
$\cn=8$ supergravity amplitudes with external scalars must obey the low energy theorems of the spontaneously broken $ E_{7(7)}$ symmetry of the theory.  These theorems require that the single soft scalar limits of any amplitude vanish, and this becomes a test for the matrix elements of candidate counterterms for the theory. If a supersymmetric counterterm is \emph{not} $E_{7(7)}$-compatible, its matrix elements have single soft scalar limits that can be collected into $SU(4)\!\times\! SU(4)$ invariant superamplitudes, as we now explain.

A scalar such as
    $\phi^{1234}$ transforms in the {\bf 70} representation of $SU(8)$, but it is a singlet of the $SU(4)$ subgroup which acts on the chosen set of indices $1234$ or on the complementary set $5678$.  Hence every scalar is invariant under a particular  $SU(4)\!\times\! SU(4)$ subgroup of $SU(8)$.
Let us consider an amplitude
 $M_n = \bigl\< \co_1\cdots \co_{n\dash1} \phi^{1234}_n \bigr\>$
containing a scalar and an unspecified set of $n\!-\!1$  particles (which may include other
scalars).  Overall $SU(8)$ invariance requires that the multiparticle state
 $\co_1\cdots \co_{n\dash 1}$ is also invariant under the $SU(4)\!\times\! SU(4)$ subgroup that preserves $1234$. The single soft scalar limit of $M_n$ gives
\be
 \lim_{p_n \to 0} M_n = C_{n-1}(\co_1\cdots \co_{n\dash1})\,.
\lab{cndef}
\ee
If non-vanishing, the amplitude $C_{n-1}$ is $SU(4)\!\times\! SU(4)$-invariant \emph{and} still subject to the SUSY Ward identities. Thus it makes sense to package  the $C_{n-1}$ in $SU(4)\!\times\! SU(4)$-invariant superamplitudes.
These transform in the ${\bf 70}$ of $SU(8)$.

 \item
Consider a toroidal compactification of string theory to four dimensions where the massless spectrum of the closed string  is that of $\cn=8$ supergravity and the open strings states are those of $\cn=4$ SYM. The symmetry group of \emph{tree level} closed string amplitudes with massless external states is the $SU(4)\!\times\! SU(4)$ inherited
from the T-duality group $SO(6,6)$.
The $SU(4)\!\times\! SU(4)$ symmetry manifests itself directly in the KLT relations given above in~(\ref{KLT}).
The amplitudes on the RHS of \reef{KLT} are each $SU(4)$-invariant, so $M_n$ is manifestly invariant under $SU(4) \!\times\! SU(4)$.
In the strict supergravity limit, $\a'\to0$, $M_n$ must preserve the full $SU(8)$.
The $\a'$-corrections, however, explicitly break  $SU(8)$ to $SU(4) \!\times\! SU(4)$ at tree level.\footnote{The $SU(4)\!\times\! SU(4)$ subgroup is the one that leaves the string dilaton and axion invariant.}
A prime example of an  $SU(8)$-violating amplitude is
 $\bigl\< - - + + \phi^{1234}\bigr\>$ which has
two pairs of opposite helicity gravitons and a single scalar
 $\phi^{1234}$. This amplitude has a leading non-vanishing contribution at order $\a'^3$. A detailed discussion of this amplitude and the symmetries of string tree
 amplitudes can be found in \cite{Elvang:2010kc}.
\end{itemize}

Consider  a general $SU(4)\!\times\! SU(4)$-invariant amplitude
$C_n(B^{\cdots}_1B^{\cdots}_2\cdots B^{\cdots}_n)$.  Each particle carries up to 8  indices of the full set $12345678$.  Suppose that the indices $1234$ and $5678$ transform under the left and right $SU(4)$ factors of the product group, respectively.  Then $SU(4)\!\times\! SU(4)$ invariance requires that each index in the set $1234$ appears $k+2$ times among the particle labels and that each  index in the set $5678$ appears $\tk +2$ times.\footnote{$SU(8)$-invariant N$^k$MHV amplitudes vanish unless the external states, labeled with upper indices, saturate an integer number of 8-index Levi-Civita tensors $(\eps^{........})^{k+2}$. Similarly, for $SU(4) \!\times\! SU(4)$ invariance, the tensor structure  $(\eps^{....})^{k+2}(\tilde{\eps}^{....})^{\tk+2}$ characterizes the N$^{(k,\tilde{k})}$MHV sector.
} This gives a natural N$^{(k,\tilde{k})}$MHV classification of such amplitudes,
 characterized by a pair of integers $(k,\tk)$. If the amplitude arises as the single soft scalar limit  \reef{cndef} of an $SU(8)$-invariant amplitude, then it has  $\tk =k -1$. We
 consider amplitudes which are characterized by two \emph{independent} integers $k$ and $\tk$; these are relevant for the analysis of closed string tree amplitudes.

 The  N$^{(k,\tilde{k})}$MHV amplitudes satisfy
 SUSY Ward identities of the \emph{same} form as $SU(8)$-invariant supergravity amplitudes.  Therefore we package all $SU(4)\!\times\! SU(4)$-invariant amplitudes in each class into superamplitudes $\cc^\text{\rm N$^{(k,\tk)}$MHV}_n$.
They are polynomials of degree $4(k+2)$ in the Grassmann variables $\h_{ia},~a=1,2,3,4$ and of degree $4(\tk +2)$ in the variables $\h_{ia},~a=5,6,7,8$.
Next we discuss the construction of the simplest of these superamplitudes.

\subsection{MHV,  ${\bf \sqNMHV}$, and N'MHV superamplitudes}
 For amplitudes in the MHV (\ie N$^{(0,0)}$MHV) sector, the SUSY Ward identities, independent of R-symmetry, determine the unique superamplitude,
 \begin{equation}\label{MHVsup}
     \mathcal{M}_n^{\rm MHV}~=~\frac{\delta^{(16)}\bigl(
     \tQd
     \bigr)}{\<n-1,n\>^8}\,\times\,\bigl\<+\cdots +--\bigr\>\,,
 \end{equation}
with a single basis element. The $\delta^{(16)}$-function is automatically $SU(8)$-invariant, as shown in
Sec.~\ref{s:Rsym},
so there are no MHV superamplitudes that violate $SU(8)$. From the point of view of KLT \reef{KLT},  each of the  MHV superamplitudes  on the open string side contain a $\d^{(8)}$-function, giving $\delta^{(8)}\times \tilde{\delta}^{(8)}=\delta^{(16)}$.

The first true $SU(4)\!\times\! SU(4)$ superamplitude sits at the N$^{(1,0)}$MHV$\oplus$N$^{(0,1)}$MHV level, which we call the
 ``{\bf$\sqrt{\text{{\bf N}}}$MHV sector}'' for simplicity. We impose a $\Z_2$-exchange symmetry between the two $SU(4)$ factors.\footnote{This is motivated by the symmetry of the closed string amplitudes we consider here under exchange of left- and right-movers.
}
Clearly, $\sqNMHV$ amplitudes break $SU(8)$.
To construct the $\sqNMHV$ superamplitude
we define
\begin{equation}\label{introdefYs}
   Y_{ijkl} ~=~
 [n\!-\!3,n\!-\! 2]^{-4}
 \times \,m_{i,n\dash 3,n\dash 2;1}\,\,m_{j,n\dash 3,n\dash2;2}\,\,m_{k,n\dash 3,n\dash 2;3}\,m_{l,n\dash 3,n\dash 2;4}\,\,\,.
\end{equation}
 Full symmetrization of its indices (see \reef{Xs})
 makes $Y_{(ijkl)}$
 invariant under $SU(4)\!\times \!SU(4)$.
We also need  the analogous polynomial $\widetilde{Y}_{(ijkl)}$ that
 depends on the $\eta_{ia}$ with $a=5,6,7,8$.
The $n$-point  $\sqNMHV$ superamplitude then takes the form
\begin{equation}\label{sqNMHV}
    {\cal M}_n^\sqNMHV~=~\frac{\delta^{(16)}\bigl(
    \tQd
    \bigr)}{\<n-1,n\>^8 }\times\!\!\!
    \sum_{1\leq i\leq j\leq k\leq l\leq n-4}\!\!\!M_n(\{i,j,k,l\}++--) \,\,\Bigl[Y_{(ijkl)}+
    \widetilde{Y}_{(ijkl)}\Bigr]\,.
\end{equation}
Note that $Y$ and $\widetilde{Y}$ are multiplied by the same basis amplitudes due to the $\Z_2$-exchange symmetry.
The basis amplitudes $M_n(\{i,j,k,l\}++--)$ have the indicated gravitons on the last four lines. Their particle content on the remaining lines is determined by the set $\{i,j,k,l\}$, which
indicates that state $i$ carries $SU(4)$ index 1, state $j$ carries $SU(4)$ index 2, and so on.  For example:
\begin{equation}
    M_6(\{1,1,1,1\}++--)\,\equiv\,\bigl\<\phi^{1234}+++--\bigr\>\,, \quad M_6(\{1,1,2,2\}++--)\,\equiv\,\bigl\<v^{12}v^{34}++--\bigr\>\,.
\end{equation}

In general, there are ${n-1 \choose 4}$ basis amplitudes at the $\sqNMHV$ level.
At the 5-point level, there is precisely  one  basis amplitude, namely $M_5(\{1,1,1,1\}++--)$, and the superamplitude
is given by
\begin{equation}
\label{sqNM5}
    {\cal M}_5^\sqNMHV
    =\frac{\delta^{(16)}\bigl(
    \tQd
    \bigr)}{\<45\>^8}\,\bigl[Y_{1111}+\widetilde{Y}_{1111}\bigr]\times \bigl\<\phi^{1234}++--\bigr\>\,.
\end{equation}

The next $SU(8)$-violating sector is N$^{(2,0)}$MHV$\oplus$N$^{(0,2)}$MHV sector, which we denote {\bf N$'$MHV} for brevity. For $n=6$, there is only one basis amplitude,\footnote{For $n=6$, this sector occurs in KLT from anti-MHV$\times$MHV open string factors.} $M_6(\{1,1,1,1\},\{2,2,2,2\}++--)$
and the superamplitude is given by
\begin{equation}\label{NpMHV}
    {\cal M}_6^{\rm N'MHV}~=~ \frac{\delta^{(16)}\bigl(
    \tQd
    \bigr)}{\<n-1,n\>^8}
    \,\Bigl[Y_{1111}Y_{2222}+\widetilde{Y}_{1111}\widetilde{Y}_{2222}\Bigr]\times
    \bigl\<\phi^{1234}\phi^{1234}++--\bigr\>\,.
\end{equation}
The superamplitudes \reef{sqNM5} and \reef{NpMHV} manifestly violate $SU(8)$ and thus vanish
in $\cn=8$ supergravity.

\subsection{The NMHV sector}
\label{s:NMHV44}
The external particles of amplitudes in the N$^{(1,1)}$MHV (=NMHV) sector are exactly as in the NMHV amplitudes we studied in
Sec.~\ref{secNMHV}.
The NMHV sector therefore includes both $SU(8)$- and $SU(4)\!\times\! SU(4)$-invariant superamplitudes. Amplitudes with particles of $SU(8)$-equivalent labels, such as
\bea
  \bigl\<\phi^{1234}\,\phi^{5678}\,++--\bigr\>\,, ~~~~
  \bigl\<\phi^{12|56}\phi^{34|78}++--\bigr\>\,, ~~~~
  \bigl\<\phi^{123|8}\phi^{4|567}++--\bigr\>\,,
\eea
must be identical if $SU(8)$-invariance is imposed, but they can be distinct in the case of $SU(4)\!\times\! SU(4)$ symmetry. The $SU(8)$-singlet  NMHV superamplitudes were given in \reef{repgrav}.  It should be contrasted with the  more general   $SU(4)\!\times\! SU(4)$ NMHV superamplitude
which  takes the form
\begin{equation}\label{CsqNMHV}
    {\cal M}_n^\text{NMHV}~=~\frac{\delta^{(16)}\bigl(
    \tQd
    \bigr)}{\<n-1,n\>^8 }\!\!\!
    \sum_{{}^{1\leq i\leq j\leq k\leq l\leq n-4}_{1\leq p\leq q\leq u\leq v \leq n-4}}\!\!\!M_n({\{i,j,k,l|p,q,u,v\}};++--) \,Y_{(ijkl)}\widetilde{Y}_{(pquv)}\, .
\end{equation}
The set $\{i,j,k,l|p,q,u,v\}$ denotes the lines on which the $SU(4)\!\times\! SU(4)$ indices $1,2,3,4$ and $5,6,7,8$ are distributed. For example,
\begin{equation}
    M_6({\{1,1,2,2|1,2,2,2\}};++--)~=~\bigl\<\chi^{12|5}\chi^{34|678}++--\bigr\>\,.
\end{equation}
If, in addition, the $\Z_2$ symmetry is imposed, we have
\begin{equation}
    M_n({\{i,j,k,l|p,q,u,v\}};++--)~=~M_n({\{p,q,u,v|i,j,k,l\}};++--)\,.
\end{equation}

Due to the reduced constraints from R-symmetry, more basis amplitudes are required for $SU(4)\!\times\! SU(4)$ NMHV superamplitudes \reef{CsqNMHV} than for $SU(8)$-invariant ones \reef{repgrav}. For example, the algebraic basis for the 6-point $SU(4)\!\times\! SU(4)$ NMHV superamplitude, with $\Z_2$ symmetry, contains 15 basis amplitudes, whereas only
9 were needed with full $SU(8)$ R-symmetry. There are {\em functional} dependencies among the 15 basis amplitudes, because lines $1$ and $2$ can be exchanged due to permutation symmetry. Taking functional relations between basis amplitudes into account, we obtain a functional basis
with $SU(4)\!\times\! SU(4)\!\times\! \Z_2$ symmetry. It consists of the following $9$ amplitudes:
\begin{equation}\label{bs}
\begin{split}
    &\bigl\<+-++-\,-\bigr\>\,, ~  \qquad \qquad \quad \bigl\<\psi^1\psi^{234|5678}++--\bigr\>\,,\qquad~ \bigl\<v^{12}v^{34|5678}++--\bigr\>\,,\\
    &\bigl\<v^{1|8}v^{234|567}++--\bigr\>\,,\qquad  \bigl\<\chi^{123}\chi^{4|5678}++--\bigr\>\,, \qquad~ \bigl\<\chi^{12|5}\chi^{34|678}++--\bigr\>\,,\\
    &\bigl\<\phi^{1234}\phi^{5678}++--\bigr\>\,,\qquad \, \bigl\<\phi^{123|8}\phi^{4|567}++--\bigr\>\,, \qquad ~\bigl\<\phi^{12|56}\phi^{34|78}++--\bigr\>\,.
\end{split}
\end{equation}
Imposing full $SU(8)$ is equivalent to demanding
\begin{equation}
\begin{split}\label{SU8cond}
     \bigl\<v^{12}v^{34|5678}++--\bigr\>~&=~\bigl\<v^{1|8}v^{234|567}++--\bigr\>\,,\\
    \bigl\<\chi^{123}\chi^{4|5678}++--\bigr\>~&=~\bigl\<\chi^{12|5}\chi^{34|678}++--\bigr\>\,, \\
    \bigl\<\phi^{1234}\,\phi^{5678}\,++--\bigr\>~&=~\bigl\<\phi^{12|56}\phi^{34|78}++--\bigr\>~=~\bigl\<\phi^{123|8}\phi^{4|567}++--\bigr\>\,.
\end{split}
\end{equation}
These $SU(8)$ conditions reduce the
number of
 functional basis elements to the 5 ones of~(\ref{SGNMHV}).

\subsection{Application to closed string tree amplitudes}

We now use the
$SU(4)\!\times\!SU(4)$ superamplitudes to
describe tree-level closed string amplitudes in toroidal compactification to four dimensions.
The symmetry group of tree level closed string amplitudes with massless external states is $SU(4)\!\times\! SU(4)\!\times\!\Z_2$, where the $\Z_2$-symmetry exchanges the $L$ and $R$-movers.

The open-string amplitudes on the right-hand side of the KLT relations \reef{KLT} are each $SU(4)$-invariant, so $M_n$ is manifestly invariant under $SU(4) \!\times\! SU(4)$. In the strict supergravity limit, $\a'\to0$, $M_n$ must preserve the full $SU(8)$, but the $\a'$-corrections explicitly break  $SU(8)$ to $SU(4) \!\times\! SU(4)$.  As discussed above, MHV amplitudes preserve the full $SU(8)$ symmetry, so the simplest possible $SU(8)$-violating amplitude belongs to the 5-point $\sqNMHV$ sector. Expanding the closed string amplitude in small $\alpha'$, we find
\bea
  \label{sqN5pt}
   \bigl\< \phi^{1234} + + - - \bigr\>_\text{closed}
   ~=~ 6 \,\zeta(3)\, \alpha'^3\, [23]^4 \<45\>^4 + O(\a'^5) \, .
\eea
This basis amplitude alone determines the full 5-point $\sqNMHV$ superamplitude  \reef{sqNM5}.

The 6-point N$'$MHV superamplitude \reef{NpMHV} is also determined by a single basis amplitude: expanding the closed string amplitude we find
\bea
  \label{Np6pt}
    \bigl\<\phi^{1234}\phi^{1234}++--\bigr\>_\text{closed}
    = - 24 \,\zeta(3)\, \alpha'^3\, [34]^4 \<56\>^4 + O(\a'^5) \, .
\eea
{From} the point of view of the string effective action \reef{effS}, the origin of the leading results for the two $SU(8)$-violating amplitudes \reef{sqN5pt} and \reef{Np6pt} is the operator
\bea
  \label{effR4}
   -2 \zeta(3)\,\a'^3 e^{-6\phi} R^4 = -2 \zeta(3)\,\a'^3 (1 - 6 \phi + 36 \phi^2 + \dots) R^4 \, ,
\eea
where the dilaton is $\phi = \tfrac{1}{2}(\phi^{1234}+\phi^{5678})$. Indeed one can match \cite{Elvang:2010kc} the numerical coefficients of relevant local 4-, 5- and 6-point amplitudes to the numerical coefficients in \reef{effR4}.

Note that no pole terms contribute to the $SU(8)$-violating basis amplitudes
\reef{sqN5pt} and \reef{Np6pt}. At the $\alpha'^3$ order, Feynman pole diagrams involve a single insertion of an interaction vertex from \reef{effR4} together with vertices from the
supergravity
theory. One can show that for the $n\le6$ amplitudes we discuss, any $SU(8)$-violation comes from local interaction terms \cite{Elvang:2010kc}.

 At the NMHV level, the most general $SU(4)\times SU(4)$-invariant NMHV superamplitude is characterized by the 9 basis amplitudes given in~(\ref{bs}). The 6-point NMHV
 closed-string amplitude is not $SU(8)$ invariant, and therefore its basis amplitudes do not satisfy the constraints~(\ref{SU8cond}). However, it is possible to decompose the closed-string amplitude $\mathcal{M}_6^\text{closed}$ into an $SU(8)$-singlet piece, $\mathcal{M}_6^{\rm singlet}$, which satisfies the $SU(8)$ conditions~(\ref{SU8cond}), and a remainder piece, $\mathcal{M}_6^{SU(4)\! \times\! SU(4)}$, which transforms non-trivially under $SU(8)$:
\bea
  \label{sumM6}
  \mathcal{M}_6^\text{closed} ~=~
  \mathcal{M}_6^{\rm singlet}  + \, \mathcal{M}_6^{SU(4)\! \times\! SU(4)}\,.
\eea
Interestingly, the $SU(8)$-violating piece has to be local because no $SU(8)$-violating pole term can contribute to the $6$-point NMHV sector. One can thus use the method described in
Sec.~\ref{s:NMHV44}
 to determine the matrix elements of the general $SU(4)\!\times\! SU(4)\!\times\!\Z_2$-preserving operator at order $\alpha'^3$.
First recall from
Sec.~\ref{s:NMHV44}
that the $6$-point
NMHV superamplitude $\mathcal{M}_6^{SU(4)\!\times\!SU(4)}$ is determined by the nine basis amplitudes \reef{bs}. At order $\alpha'^3$ of the closed string amplitude, these basis amplitudes have to be local and of dimension 8 (because $R^4$ contains 8 derivatives). Little group scaling implies that only the 3 scalar basis amplitudes of \reef{bs} can be non-vanishing; indeed,
they must each be
equal to $[34]^4 \<56\>^4$ times a numerical coefficient. The ratio of the coefficients for the three scalar amplitudes is fixed by requiring that the resulting superamplitude
$\mathcal{M}_6^{SU(4)\!\times\!SU(4)}$
 is permutation invariant.
All included, this uniquely fixes the superamplitude $\mathcal{M}_6^{SU(4)\!\times\!SU(4)}$
to be
\begin{equation}
\begin{split}\label{O6NMHVsuper}
        \mathcal{M}_6^{SU(4) \!\times\! SU(4)}
        =-\tfrac{6}{5}\zeta(3)\,\alpha'^3\times\delta^{(16)}\!\bigl(
        \tQd
        \bigr)\,
        [34]^4\<56\>^{-4} \Bigl[&12\bigl(Y_{1111}\widetilde{Y}_{2222}+Y_{2222}\widetilde{Y}_{1111}\bigr)
        +2\,Y_{(1122)}\widetilde{Y}_{(1122)}\\
        &-3\bigl(Y_{(1112)}\widetilde{Y}_{(1222)}+Y_{(1222)}\widetilde{Y}_{(1112)}\bigr)\Bigr]~
        + \,O(\alpha'^5)\,
\end{split}
\end{equation}
The overall normalization is  fixed  by the explicit closed-string amplitude computation \cite{Elvang:2010kc}. This superamplitude~(\ref{O6NMHVsuper}) encodes  the $SU(8)$-{\em violating} component of $\phi^2R^4$ in the NMHV sector. One can also check directly that no $SU(8)$-singlet contribution `hides'
  in \reef{O6NMHVsuper}: if we average the superamplitude~(\ref{O6NMHVsuper}) over $SU(8)$, using the formula~(\ref{su8avg}), we find that it is  $\propto12+2\times 18 - 3 \times 16 = 0$.

The $\alpha'^3$ term in $\mathcal{M}_6^{\rm singlet}$ is precisely the superamplitude that encodes  the $6$-point NMHV matrix elements of the $SU(8)$-invariant operator $R^4$. In
Sec.~\ref{secE77},
we only computed its matrix element $\<\varphi\bar\varphi ++--\>$ to rule out $R^4$ as a potential counterterm in $\cn=8$ supergravity. However, the entire 6-point NMHV superamplitude of $R^4$ can be determined from~(\ref{sumM6}) in terms of closed-string tree amplitudes, using~(\ref{O6NMHVsuper}) to subtract off the $SU(8)$-violating contribution. The results for the single soft scalar limits agree with those obtained from the $SU(8)$-averaging procedure.

\section*{Acknowledgements}
We thank N.~Beisert, A.~Morales and S.~Stieberger for collaboration on topics reviewed in this paper.
The research of DZF is supported by NSF grant PHY-0967299 and by
the US Department of Energy through cooperative research agreement DE-FG-0205FR41360.
HE is supported by NSF CAREER Grant PHY-0953232, and in part by the US Department of Energy under DOE grants DE-FG02-95ER 40899 (Michigan) and DE-FG02-90ER40542 (IAS).
The research of MK is supported by the NSF grant PHY-0756966.

\appendix

\setcounter{equation}{0}
\section{Derivation of solution to NMHV SUSY Ward identities}
\label{app1}

We provide in this appendix the missing steps in the derivation-outline presented in
Sec.~\ref{secstrategy}.
In {\bf step 2}
of Sec.~\ref{secstrategy},
all $\eta_{n-1,a}$ and $\eta_{n,a}$ are eliminated from $P_4$. This is done by first using the Schouten identity to rewrite the Grassmann $\delta^{(8)}$-function of \reef{Namp} as
\be
 \lab{first}
  \delta^{(8)} \Big( \sum_{i=1}^n |i\> \, \eta_{ia}\Big)
 ~=~
  \frac{1}{\<n-1,n\>^4}~
  \delta^{(4)} \Big( \sum_{i=1}^n \<n-1,i\> \, \eta_{ia}\Big)~
  \delta^{(4)} \Big( \sum_{j=1}^n \<n j\> \, \eta_{ja}\Big) \,,
\ee
The two $\delta^{(4)}$-functions can be used to express $\eta_{n-1,a}$ and $\eta_{na}$ in terms of the other $\eta_{ia}$'s, specifically
\bea\label{etaelim}
  \eta_{n-1, a} = -
  \sum_{i=1}^{n-2} \frac{\< n i \>}{\< n,n-1 \>} \, \eta_{i a}
  \, ,~~~~~
  \eta_{n a} =
  - \sum_{i=1}^{n-2} \frac{\< n-1, i \>}{\< n-1,n \>} \, \eta_{i a} \, .
\eea
Inserting this into the $P_4$ of \reef{Namp}, we find
\be
  \lab{step2}
 P_4 = \frac{1}{\<n-1,n\>^4} \sum_{i,j,k,l=1}^{n-2} c_{i j k l}\,
 \eta_{i 1} \, \eta_{j 2} \,  \eta_{k 3} \,  \eta_{l 4} \, .
\ee
The $c_{ijkl}$'s are linear combinations of the $q_{ijkl}$'s, but we will not need their detailed relationship.

In {\bf step 3} of
Sec.~\ref{secstrategy},
we presented the solution to the $Q^a$-Ward identities $Q^a\,P_4 = 0$. The action of $Q^1$ on $P_4$ gives
\be
  0 ~=~ Q^1 P_4
     ~\propto~ \sum_{i,j,k,l=1}^{n-2} [\eps i ] \, c_{ijkl}\,
      \eta_{j2}\,\eta_{k3}\,\eta_{l4}
     ~=~
     \sum_{j,k,l=1}^{n-2} \Big[ \sum_{i=1}^{n-2} [\eps i ] \, c_{ijkl}  \Big]
      \eta_{j2}\,\eta_{k3}\,\eta_{l4}\, .
\ee
The quantity in square brackets must vanish for any triple $jkl$, so the $c_{ijkl}$ must satisfy
\be
  \lab{QWI}
  \sum_{i=1}^{n-2}\, [\eps i ] \, c_{ijkl}  ~=~ 0 \, .
\ee
This is the relation quoted in {\bf step 3} of
 Sec.~\ref{secstrategy}.

Now select two arbitrary (but fixed) lines $s$ and $t$ among the remaining lines $1,\ldots, n-2$.
We choose the SUSY spinor $|\e] \sim |t]$ and then  $|\e] \sim |s]$  and  use \reef{QWI} to
express the coefficients $c_{sjkl}$ and $c_{tjkl}$ in terms of $c_{ijkl}$ with
$i \neq s,t$:
\bea \lab{c1c2}
    c_{sjkl} = - \sum_{i\neq s,t}^{n-2} \frac{[t i ]}{[ts]} \, c_{ijkl}\, ,
    \hspace{5mm}
    c_{tjkl} = - \sum_{i\neq s,t}^{n-2} \frac{[s i ]}{[st]} \, c_{ijkl} \, .
\eea
The sums extend from $i=1$ to $i=n-2$, excluding lines $s$ and $t$.
We can write similar relations for $c_{iskl}, ~c_{itkl}$, etc.~using supercharges $Q^a$, $a=2,3,4$\,, in the same way.  We use the relations~(\ref{c1c2}) to write $P_4$ in \reef{step2} as
 \bea
 \nonumber
 \<n-1,n\>^4 P_4 &=& \sum_{\,j,k,l = 1}^{n-2} \,\sum_{i\neq s,t}^{n-2} c_{ijkl}\,\,\eta_{i1} \eta_{j2}\,\eta_{k3}\,\eta_{l4} +\sum_{j,k,l=1}^{n-2}\big(
 c_{sjkl}\,\eta_{s1} + c_{tjkl}\,\eta_{t1}\big)\,\eta_{j2}\,\eta_{k3}\,\eta_{l4}\\
 &=& \frac{1}{[st]} \sum_{\,j,k,l = 1}^{n-2} \,\sum_{i\neq s,t}^{n-2} c_{ijkl}\, \pdan_{ist,1} \,\eta_{j2}\,\eta_{k3}\,\eta_{l4}\,,
 \lab{gdstuff}
 \eea
in which $\pdan_{ist,1}$ is the first-order polynomial
introduced in \reef{pdanIntro}.   We repeat this process and use the analogues of \reef{c1c2} for
$c_{iskl}$ and $c_{itkl}$ to reexpress the
 sum over $j$ in \reef{gdstuff} in terms of $\pdan_{jst,2}$.
Repeating
this
for the $k$ and $l$ sums, we find
\be
  \lab{almostdone}
  \ca_n^\text{NMHV} ~= \sum_{i,j,k,l\neq s,t}^{n-2} c_{ijkl} \,X_{ijkl}\,,
  ~~~~~~~~
  X_{ijkl} ~\equiv~ \d^{(8)}\big(\tQ_a\big)
  ~ \frac{m_{i s t,1}\,m_{j s t,2}\,m_{k s t,3}\,m_{l s t,4}}{[s t]^4\<n-1,n\>^4}\,.
\ee
The $\eta$-polynomial $X_{ijkl}$ of degree $12$ in  \reef{almostdone} are
manifestly invariant under both $\widetilde{Q}_a$ and $Q^a$ supersymmetry.
As in
 Sec.~\ref{secstrategy},
 it is convenient to set $s=n\!-\!3$ and $t=n\!-\!2$.
Since the $c$-coefficients are fully symmetric we can symmetrize the $X$-polynomials and write
\be
   \lab{step35}
   \ca_n^\text{NMHV} ~=
     \sum_{1\le i\le j\le k\le l \le n-4}
     c_{ijkl}\,     X_{(ijkl)}   \, ,\hspace{1.3cm}
  X_{(ijkl)} \equiv \sum_{\mathcal{P}(i,j,k,l)} X_{ijkl} \, .
\ee
The sum
 over permutations $\mathcal{P}(i,j,k,l)$ in the definition of $X_{(ijkl)}$ is explained below \reef{defXs}.

In the final {\bf step 4} we identify the coefficients $c_{ijkl}$ as on-shell amplitudes of the basis. Recall \cite{Bianchi:2008pu} that component amplitudes are obtained by applying Grassmann derivatives to the superamplitude.
Consider amplitudes with negative-helicity gluons at positions $n\!-\!1$ and $n$. To extract such amplitudes from \reef{step35} we apply four $\eta_{n-1,a}$-derivatives and four $\eta_{na}$-derivatives to $\ca_n^\text{NMHV}$. These derivatives must hit the Grassmann $\d$-function and the result is simply a factor $\< n-1,n\>^4$, which cancels the same factor in the denominator of $X_{(ijkl)}$.
We must apply four more Grassmann derivatives
 $\frac{\pa}{\pa \eta_{i1}}\frac{\pa}{\pa \eta_{j2}}\frac{\pa}{\pa \eta_{k3}}\frac{\pa}{\pa \eta_{l4}}$ to $\ca_n$ in order to extract an NMHV amplitude. These derivatives hit the product of $\pdan_{i,n\dash3,n\dash2;a}$-polynomials and produce a factor of $[n\!-\!3,n\!-\!2]^4$ which cancels the
 remaining denominator factor of $X_{ijkl}$. As a result, the 12 $\eta$-derivatives just leave us with the coefficient $c_{ijkl}$. When
$\!1\!\le\! i\! \le\! j\! \le\! k\! \le\! l \!\le\! n\!-\!4$\,,
we have therefore identified $c_{ijkl}$ as the amplitude
$c_{ijkl} =A_n \big( \{i,j,k,l\}++ -- \big)$. As discussed in the main text, the notation means that line $i$ carries the $SU(4)_R$ index 1, line $j$ carries index 2 etc.
With this identification of the $c_{ijkl}$ coefficients we can now write the final result \reef{step4} in terms of the
\mbox{$(n-4)(n-3)(n-2)(n-1)/4! = {n-1 \choose 4}$}
 basis amplitudes $A_n \big( \{i,j,k,l\}++ -- \big)$ of the algebraic basis.

\end{document}